\documentclass[preprint,10pt]{elsarticle}

\usepackage{hyperref} 
\hypersetup{ 
  colorlinks   = true, 
  urlcolor     = blue, 
  linkcolor    = blue, 
  citecolor    = blue, 
} 
\usepackage[fleqn]{amsmath}
\usepackage{amsfonts}
\usepackage{amssymb}
\usepackage{amsthm}

\usepackage{pifont}
\usepackage{natbib}
\usepackage{geometry}
\usepackage{graphicx}
\usepackage{txfonts}

\usepackage[utf8]{inputenc}
\usepackage{caption}
\usepackage{subcaption}
\usepackage[usenames, dvipsnames]{color}

\journal{International Journal of Heat and Mass Transfer}

\biboptions{sort&compress}

\begin{document}

\begin{frontmatter}

\title{Two-phase developing laminar mixing layer at supercritical pressures}

\author{Branson W. Davis\fnref{myfootnote1}\corref{mycorrespondingauthor}}
\ead{davisbw@uci.edu}
\author{Jordi Poblador-Ibanez\fnref{myfootnote2}}
\author{William A. Sirignano\fnref{myfootnote3}}
\address{University of California, Irvine, CA 92697-3975, United States}
\fntext[myfootnote1]{Undergraduate Student Researcher, Department of Mechanical and Aerospace Engineering.}
\fntext[myfootnote2]{Graduate Student Researcher, Department of Mechanical and Aerospace Engineering.}
\fntext[myfootnote3]{Professor, Department of Mechanical and Aerospace Engineering.}


\cortext[mycorrespondingauthor]{Corresponding author}


\begin{abstract}

Numerical analysis of a shear layer between a cool liquid \(n\)-decane hydrocarbon and a hot oxygen gas at supercritical pressures shows that a well-defined phase equilibrium can be established. Variable properties are considered with the product \(\rho \mu\) in the gas phase showing a nearly constant result within the laminar flow region with no instabilities. Sufficiently thick diffusion layers form around the liquid-gas interface to support the case of continuum theory and phase equilibrium. While molecules are exchanged for both species at all pressures, net mass flux across the interface shifts as pressure is increased. Net vaporization occurs for low pressures while net condensation occurs at higher pressures. For a mixture of \(n\)-decane and oxygen, the transition occurs around 50 bar. The equilibrium values at the interface quickly reach their downstream asymptotes. For all cases, profiles of diffusing-advecting quantities collapse to a similar solution (i.e., function of one independent variable). Validity of the boundary layer approximation and similarity are shown in both phases for Reynolds numbers greater than 239 at 150 bar. Results for other pressures are also taken at high Reynolds numbers. Thereby, the validity of the boundary layer approximation and similarity are expected. However, at very high pressures, the similar one-dimensional profiles vary for different problem constraints.

\end{abstract}

\begin{keyword}
real-fluid laminar mixing layer \sep phase equilibrium \sep supercritical pressure \sep phase change \sep diffusion layer
\end{keyword}

\end{frontmatter}


\setlength\abovedisplayshortskip{0pt}
\setlength\belowdisplayshortskip{-5pt}
\setlength\abovedisplayskip{-5pt}

\section{Introduction}
\label{sec:intro}

A mixing layer develops between two parallel streams where there is a difference in velocity. Mixing layers are of paramount importance holistically for understanding combustion and development of turbulence in aerodynamics as well as atmospheric and oceanic flows. Chapman numerically studied the wake formed behind a body passing through air \cite{chapman1949laminar}. Specifically, the velocity profile evolution of a laminar, single-phase air mixing layer was determined as it developed behind the body. Compressibility effects were included. This problem closely resembles a mixing layer starting with zero thickness (i.e., the two flows are separated by a splitter plate initially). A similar solution to the equations of motion existed; that is, the flow field could be determined in terms of a single space variable that was a function of both physical coordinates. However, many fluid mixing layers are not single-phase, but are instead multiphase flows. 

Two-phase mixing layers pervade a myriad of industrial applications such as filtration, spray processes, fluid-particle transport, and fuel injection for propulsion systems. Thus, it is important to understand the development of such mixing layers. For a two-phase flow, mixing of species causes variable fluid properties to exist. A velocity gradient between the two streams causes a shear instability at the liquid-gas interface. As the interfacial instability grows, the bulk liquid breaks into small droplets, forming a two-phase mixing layer. This process is often referred to as "atomization". The liquid and gas streams are immiscible. Thus, the developed mixing layer consists of two distinct phases, liquid droplets formed from the bulk liquid and gas. Development of two-phase mixing layers at subcritical pressures are well understood. Doughty and Pruess numerically studied the mixing layer between water and air in a porous medium at subcritical pressures \cite{doughty1992similarity}. They exposed the flow to a nearby linear heat source and found that the partial differential equations reduced to a similarity solution. Other cases of similarity solutions are well established for both single- and two-phase flows in other works and textbooks \cite{sadatomi1982twophase,kleinstreuer2003two,white2006viscous, williams2018combustion}. 

Liquid injectors used in combustion devices are designed to optimize atomization to allow for the combustion reaction to occur. In many cases, the operating pressure can be larger than the critical pressure of the injected liquid. In this supercritical environment, the thermodynamics and fluid dynamics during injection are modified considerably. When a liquid jet is injected at a pressure or temperature higher than that of the critical point of the substance, the liquid can no longer considered as an incompressible fluid with high surface tension \cite{hirschfelder1964molecular,prausnitz1998molecular}. It no longer behaves like a traditional two-phase mixing layer \cite{chehroudi1999initial, maslowe1971inviscid, mayer1998propellant}. The jet changes to resemble more a turbulent, gaseous jet. In addition, there is no longer evidence of droplet formation. Instead, thread-like geometries emerge from the jet which dissolve away from the jet core \cite{chehroudi1999initial}. Experimental studies have shown that a thermodynamic transition occurs where the liquid and gas exhibit similar fluid properties \cite{hsieh1991droplet,delplanque1993numerical, yang1994vaporization,sirignano1997selected,juanos2015thermodynamic}. Understanding how the shear layer between the liquid and gas evolves at supercritical pressures is crucial to understanding the initial stages of high-pressure atomization. 

There have been many experimental investigations of supercritical phenomena \cite{mayer1996propellant,mayer2000injection,segal2008subcritical,chehroudi2012recent}, but they are limited by measurement techniques and high costs associated with the extreme environments. Computational modeling allows for more accurate simulation of these environments and complements experimental data. Past works assumed a two-phase behavior could not be maintained under the altered thermodynamics \cite{spalding1959theory,rosner1967liquid}. However, many studies have found a two-phase behavior contingent on thermodynamic phase equilibrium at the liquid-gas interface. Phase equilibrium enhances the dissolution of gas into the liquid phase creating substantially thick layers (i.e., of the order of micrometers) in both phases \cite{poblador2018transient}. Mixture critical properties near the interface differ from the bulk fluid critical properties and often, mixture pressures exceed chamber pressures. Because sufficiently thick diffusion layers occur in a short period, the interface can be treated as a discontinuity with a jump in fluid properties across it.

In prior studies, a similarity solution has not been found for a two-phase mixing layer at supercritical pressures with real-fluid thermodynamic modeling. Poblador-Ibanez and Sirignano suggested a self-similar behavior of a temporal one-dimensional configuration at supercritical pressures \cite{poblador2018transient}. This paper discusses the similarity in solutions of the mixing layer partial differential equations as downstream positions vary. A parallel study by Poblador-Ibanez et al. develops the similar solution by formulating and solving the appropriate ordinary differential equations \cite{poblador2020self}. In addition, the existence of a sharp phase interface is determined at pressures above the critical pressure of oxygen and \(n\)-decane with the establishment of phase equilibrium in a sufficiently short distance before a transition to turbulent flow and before hydrodynamic instabilities dominate.

\section{Two-phase laminar mixing layer}
\label{sec:mixinglayer}

\subsection{Problem definition}
\label{subsec:probdef}
\begin{figure}[h!]
\includegraphics[scale=0.55]{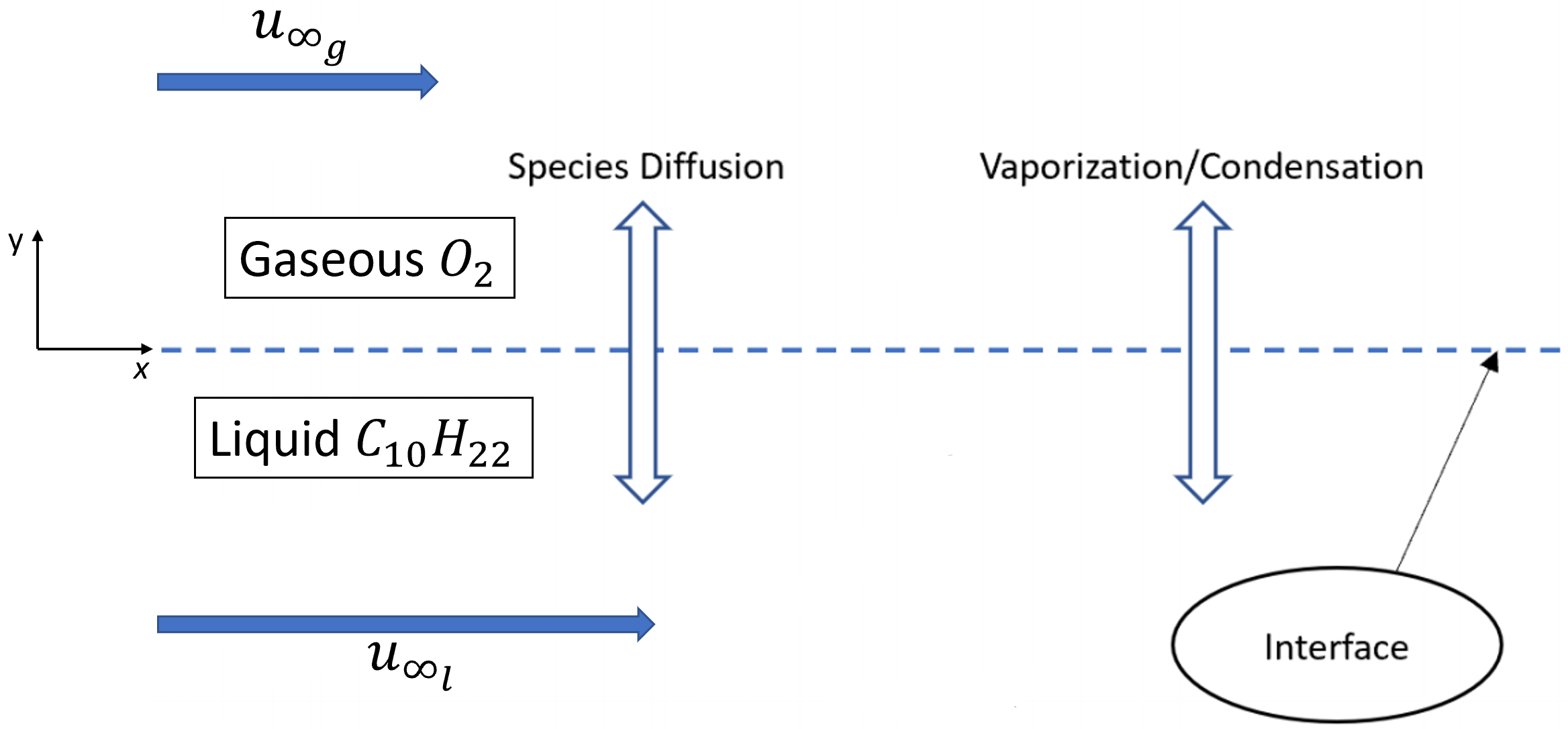}
\centering
\caption{2D schematic of the mixing layer problem.}
\label{fig:2DSchematic}
\end{figure}

As shown in Figure \ref{fig:2DSchematic}, pure liquid \(C_{10}H_{22}\) with velocity \(u_{\infty_{l}}\) is introduced into gaseous \(O_2\) with a slower velocity \(u_{\infty_{g}}\). Steady state is assumed marching downstream creating a two-dimensional problem. Transverse momentum imbalance at the edges of the developing diffusion layers becomes negligible quickly. While the interface will generally not remain at the initial y-value, it is reasonable to assume a fixed interface at \(y\) = 0 as shown in Section \ref{subsec:fixedint}. The liquid temperature remains lower than the gas temperature, within the critical temperature range for the pure liquid species. Pressure is held constant throughout the domain. Supercritical pressures enhance the dissolution of lighter gaseous \(O_2\) into the liquid phase through the imposition of thermodynamic phase equilibrium. Thus, a binary mixture exists on either side of the interface as molecules of liquid \(n\)-decane mix with the surrounding gas while gaseous molecules of \(O_2\) enter the liquid phase. A net condensation or vaporization results about the interface depending on the interface energy balance and pressure regime \cite{poblador2018transient,poblador2019analysis}.

\subsection{Laminar flow conditions and instability analysis}
\label{subsec:flowcond}
This analysis is intended to determine whether a distinct two-phase flow will be established in a short distance before hydrodynamic instabilities related to transitional turbulence appear. If that occurs, the transitional turbulence and associated vorticity dynamics becomes an essential feature in the atomization process \cite{jarrahbashi2014vorticity,jarrahbashi2016early,zandian2017planar,2017arXiv170603742Z}. A Reynolds number will be chosen such that the resulting flow is laminar and stable. Very little is known about the critical Reynolds numbers at which a two-phase laminar mixing layer becomes unstable. Even less is known about the transition Reynolds number to a turbulent flow. Huang and Ho \cite{huang1990small-scale} found that the transition displacement-thickness Reynolds number for a plane mixing layer occurred in the range \(114 < Re_{\delta^{*}} < 140\). However, the transition was not definite, but rather local Reynolds number dependent. Thus, a boundary layer approximation is assumed to provide some guidance. Tani \cite{tani1969boundary} states that two-dimensional instability oscillations begin above a displacement thickness Reynolds number \(Re_{\delta^{*}} > 450\). Taking into account a degree of uncertainty, a Reynolds number with respect to displacement thickness, \(Re_{\delta} = 100 \) (\(Re_x = \) 10,000) is used in this work. 

A streamwise domain length of 1 cm with fluid velocities of \(\mathcal{O}\) (10 m/s) were chosen to ensure a fully-developed flow while keeping the laminar and stable flow assumptions valid. To satisfy this requirement with \(Re_{\delta} = 100 \), \(\Delta u\) is computed between the free-stream liquid and gas phases using the Reynolds number definition with respect to streamwise distance, \(x\), as

\begin{equation}
Re_{x} = \frac{\rho_{\infty_{g}} x \Delta u}{\mu_{\infty_{g}}}
\label{eqn:reynolds}
\end{equation}
where the density, \(\rho_{\infty_{g}}\), and the viscosity, \(\mu_{\infty_{g}}\), are computed from the bulk gas. 

Assuming a mean flow velocity, \(u_{mean}\) = 10 m/s, streamwise velocity freestream boundary conditions, \(u_{\infty_{g}}\) and \(u_{\infty_{l}}\), are found for each pressure case. Their values are illustrated in Table \ref{tab:flowcond}. Viscosity, density, and surface tension computed at the interface are shown in Table \ref{tab:interfacecond}. More information on how to evaluate these properties is provided in Section \ref{subsec:transp}.

\begin{table}[h!]
\centering
\begin{tabular}{c c c c c}
\hline
  & p = 10 bar & p = 50 bar & p = 100 bar & p = 150 bar \\
\hline
\(u_{\infty_{g}}\) (m/s) & 7.673 & 9.525 & 9.755 & 9.830\\
\(u_{\infty_{l}}\) (m/s) & 12.327 & 10.475 & 10.246 & 10.170\\
\hline
\end{tabular}
\caption{Streamwise velocity freestream values to satisfy laminar flow conditions for each analyzed pressure.}
\label{tab:flowcond}
\end{table}

\begin{table}[h!]
\centering
\begin{tabular}{c c c c c}
\hline
  & p = 10 bar & p = 50 bar & p = 100 bar & p = 150 bar \\
\hline
\(\rho_{g}\) (kg/\(\mbox{m}^{3}\)) & 12.413 & 47.932 & 91.434 & 134.362\\
\(\rho_{l}\) (kg/\(\mbox{m}^{3}\)) & 593.529 & 580.403 & 571.833 & 563.888\\
\(\mu_{g}\) \bigg(\(\frac{\mbox{kg}}{\mbox{m}\cdot \mbox{s}}\)\bigg) & 1.799\(\times 10^{-5}\) & 2.500\(\times 10^{-5}\) & 2.661\(\times 10^{-5}\) & 2.778\(\times 10^{-5}\)\\
\(\mu_{l}\) \bigg(\(\frac{\mbox{kg}}{\mbox{m}\cdot \mbox{s}}\)\bigg) & 1.953\(\times 10^{-4}\) & 1.290\(\times 10^{-4}\) & 9.725\(\times 10^{-5}\) & 7.793\(\times 10^{-5}\)\\
\(\sigma\) (kg/\(\mbox{s}^{2}\))  & 9.705\(\times 10^{-3}\) & 7.186\(\times 10^{-3}\) & 5.075\(\times 10^{-3}\) & 3.466\(\times 10^{-3}\)\\
\hline
\end{tabular}
\caption{Interface thermodynamic conditions at x = 0.01 m for each analyzed pressure.}
\label{tab:interfacecond}
\end{table}

Under such flow conditions, it is important to analyze Kelvin-Helmholtz (KH) hydrodynamic instabilities to determine the potential effects different high-pressure environments have on surface perturbations and confirm the stability of the chosen Reynolds number. KH wave growth is evaluated within the streamwise domain at the interface, where the largest gradients exist. Small perturbations at the liquid-gas interface for liquid sheets flowing parallel to a gas can be analyzed by a linear temporal instability study \cite{rangel1991linear,joseph2007potential}. The evolution of the perturbation of the interface displacement is given by

\begin{equation}
\label{eqn:kh1}
\Omega (x,t) = \hat{\Omega} e^{\epsilon t} e^{ikx}
\end{equation} 
where \(\Omega\) represents the perturbation amplitude as a function of the initial oscillation amplitude, \(\hat{\Omega}\), growth rate, \(\epsilon\), time, \(t\), wave number, \(k=2\pi/\lambda\), and location, \(x\).

The growth rate parameter, \(\epsilon\), can be expressed as 

\begin{equation}
\label{eqn:growth}
\epsilon = \epsilon_{R} + \epsilon_{I}i
\end{equation}
where the real part, \(\epsilon_{R}\), can be analyzed to characterize the stability of a perturbation (i.e., \(\epsilon_{R} < 0\) is stable and \(\epsilon_{R} > 0\) is unstable). A linear analysis of small-amplitude interface perturbations provides an expression for \(\epsilon\) as a function of the fluid properties without the effects of gravity \cite{joseph2007potential} 

\begin{equation}
\label{eqn:kh2}
\begin{split}
\epsilon = & -i\frac{k(\rho_{g} u_{\infty_{g}}+\rho_{l} u_{\infty_{l}})}{\rho_{g}+\rho_{l}}-k^2\frac{\mu_{g}+\mu_{l}}{\rho_{g}+\rho_{l}}  \pm \bigg[\frac{\rho_{g} \rho_{l} k^2 (u_{\infty_{g}}-u_{\infty_{l}})^2}{(\rho_{g} + \rho_{l})^2} - \\
& -\frac{\sigma k^3}{\rho_{g} + \rho_{l}} + \frac{k^4(\mu_{g}+\mu_{l})^2}{(\rho_{g}+\rho_{l})^2} + 2ik^3 \frac{(\rho_{g}\mu_{l}-\rho_{l}\mu_{g})(u_{\infty_{g}}-u_{\infty_{l}})}{(\rho_{g}+\rho_{l})^2}\bigg]^{1/2}
\end{split}
\end{equation}
which includes the streamwise velocity, \(u\), viscosity, \(\mu\), density, \(\rho\), and the surface tension coefficient, \(\sigma\). \(u\) is taken in the bulk liquid and gas while \(\mu\), \(\rho\), and \(\sigma\) are obtained at the interface. Viscosity and surface tension calculations at the interface are discussed in Section ~\ref{subsec:transp}. This expression can only roughly represent the problem analyzed in this paper since it only considers normal viscous stress and ignores shear stress, thereby combining the vorticity to a zero-thickness interface. However, it can serve as a guide.

\begin{figure}[h!]
\includegraphics[scale=0.15]{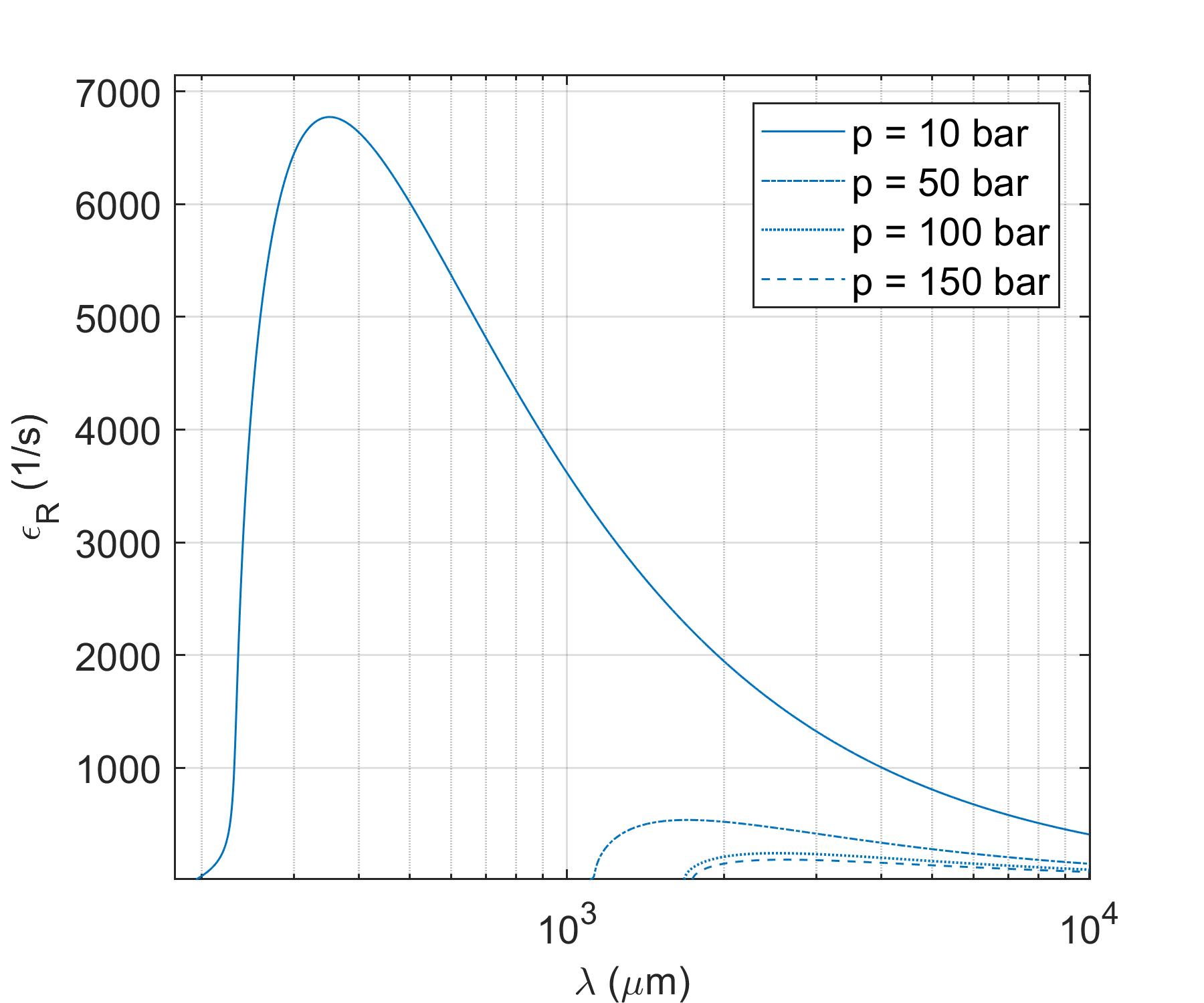}
\centering
\caption{Kelvin-Helmholtz growth rate vs. wavelength at a mean flow velocity \(u_{mean}\) = 10 m/s at \(p\) = 10, 50, 100, and 150 bar.}
\label{fig:KelvinHelm}
\end{figure}

\begin{table}[h!]
\centering
\begin{tabular}{c c c c}
\hline
\(p\) (bar) & \(\lambda_w\) (\(\mu\)m) & \(\epsilon_{\text{R}}\) (1/s) & \(\tau\) (ms) \\
\hline
 10  & 3.510\(\times 10^2\) & 6.773\(\times 10^3\) & 0.148 \\
 50  & 1.699\(\times 10^3\) & 5.377\(\times 10^2\) & 1.860\\
100  & 2.525\(\times 10^3\) & 2.428\(\times 10^2\) & 4.119\\
150  & 2.616\(\times 10^3\) & 1.854\(\times 10^2\) & 5.393\\
\hline
\end{tabular}
\caption{Kelvin-Helmholtz instability results for the oxygen/n-decane mixture.}
\label{tab:kh2}
\end{table}

Figure \ref{fig:KelvinHelm} shows the the real part of the growth rate as a function of wavelength for a mean flow velocity of \(u_{mean}\) = 10 m/s. Only \(\epsilon_{R} > 0\) is shown as it represents flow instabilities. Table \ref{tab:kh2} shows the maximum growth rate with its associated wavelength and characteristic time, \(\tau\), for the four pressure cases. For increasing pressure, there is an increase in the critical wavelength for the instabilities to develop and a decrease in the growth rate. For flow moving at 10 m/s, it takes \(\tau^*\) = 1 ms to pass through a 1 cm domain. Thus for \(p\) = 50, 100, and 150 bar, the characteristic time for instabilities to develop is larger than the time it takes the flow to pass through the domain. Therefore, instabilities are negligible while diffusion layers become sufficiently thick to use continuum theory on each side of the phase interface. 


However, note the development of instabilities within the domain for the 10 bar pressure case. As previously mentioned, Eq. (\ref{eqn:kh2}) does not take into account shear stress effects. Such effects would dampen the perturbations, thus decreasing the growth rate and moving the most unstable waves to longer wavelengths \cite{jarrahbashi2014vorticity,jarrahbashi2016early,zandian2017planar,2017arXiv170603742Z}. Consequently, it is expected that the 10 bar case would also be stable or very slowly growing in amplitude within the analyzed domain.



\subsection{Governing equations}
\label{subsec:gover}

For a sufficiently large Reynolds number, the governing equations for a two-dimensional laminar mixing layer follow the boundary-layer approximation~\cite{white2006viscous}, where \(v\ll u\), partial derivatives in \(x\) are much smaller than partial derivatives in \(y\) and \(\partial p/\partial y \approx 0\). By assuming constant pressure everywhere, \(\partial p/\partial x = \partial p/\partial y = 0\). Under these assumptions, the transverse momentum equation is replaced by the pressure assumption and the transverse velocity is directly obtained from the continuity equation \cite{poblador2018transient,he2017sharp}. Furthermore, the pressure term in the energy equation disappears and viscous dissipation and kinetic energy become negligible for low-Mach number flows at high pressures.  \par 

The steady-state, high-pressure, two-phase mixing layer global and species continuity equation, Eq.~(\ref{eqn:gc}) and (\ref{eqn:sc_cons}), can be expressed in conservative form as

\begin{equation}
\label{eqn:gc}
\frac{\partial}{\partial x}(\rho u)+\frac{\partial}{\partial y}(\rho v)=0
\end{equation}

\begin{equation}
\label{eqn:sc_cons}
\frac{\partial}{\partial x}(\rho u Y_i)+\frac{\partial}{\partial y}(\rho v Y_i ) + \frac{\partial}{\partial y}(J_{i}^{y})=0
\end{equation}

Using the conservative global and species continuity equations, the non-conservative forms of the species continuity equation, Eq.~(\ref{eqn:sc_noncons}), streamwise momentum equation, Eq.~(\ref{eqn:xmom_noncons}), and energy equation, Eq.~(\ref{eqn:ene_noncons}), can be written as

\begin{equation}
\label{eqn:sc_noncons}
\rho u\frac{\partial Y_i}{\partial x}+\rho v\frac{\partial Y_i}{\partial y} + \frac{\partial}{\partial y}(J_{i}^{y})=0
\end{equation}

\begin{equation}
\label{eqn:xmom_noncons}
\rho u\frac{\partial u}{\partial x} + \rho v \frac{\partial u}{\partial y}=\frac{\partial}{\partial y}\bigg(\mu \frac{\partial u}{\partial y}\bigg)
\end{equation}

\begin{equation}
\label{eqn:ene_noncons}
\rho u\frac{\partial h}{\partial x}+\rho v\frac{\partial h}{\partial y} - \rho u \sum_{i=1}^{N} h_i \frac{\partial Y_i}{\partial x} - \rho v \sum_{i=1}^{N} h_i \frac{\partial Y_i}{\partial y} + \sum_{i=1}^{N} J_{i}^{y} \frac{\partial h_i}{\partial y} = \frac{\partial}{\partial y}\bigg(\lambda \frac{\partial T}{\partial y}\bigg)
\end{equation}

\noindent
where \(\rho\), \(T\), \(u\), and \(v\) are the mixture density, temperature, streamwise velocity, and transverse velocity, respectively. \(h\) is the mixture specific enthalpy and \(Y_i\) is the mass fraction of a single mixture component where the subscript, \(i\), denotes the species. Other mixture fluid properties include the dynamic viscosity, \(\mu\), the diffusion mass flux in the transverse direction, \(J_i^y\), and the thermal conductivity, \(\lambda\). Note \(h \neq \sum^{N}_{i = 1}Y_{i}h_{i}\) because of intermolecular forces.

In the present work, thermal effects on mass diffusion are neglected and, for a binary mixture, the Maxwell-Stefan equations are solved and recast in Fickian form in a mass-based frame of reference~\cite{leahy2007unified,mutoru2011form} as:

\begin{equation}
\label{eqn:fickian}
\vec{J}_1 = -\vec{J}_2 = -\rho D_m \nabla Y_1
\end{equation}

\noindent
where \(D_m\) is the mass-based diffusion coefficient.

Freestream composition consists of pure \(n\)-decane and oxygen gas in the liquid phase and gas phase, respectively. Along with streamwise velocities in Table \ref{tab:flowcond}, freestream temperatures are 450 K and 550 K in the liquid and gas phases, respectively. Interface matching conditions are presented in Section \ref{subsec:pheq}. At x = 0, pure \(n\)-decane and oxygen exist in each phase. Likewise, a uniform temperature distribution is assumed in each phase. To ensure numerical stability, the initial streamwise velocity distribution follows a distribution resembling a hyperbolic tangent function with a thickness of 10 nodes. 
%


\subsection{Stationary interface analysis}
\label{subsec:fixedint}

A velocity correction is required to balance the transverse momentum on either side of the interface according to:

\begin{equation}
\rho_{\infty_{l}}(v_{\infty_{l}}+C)^{2}=\rho_{\infty_{g}}(v_{\infty_{g}}+C)^{2}
\label{eqn:TranMomBal}
\end{equation}

\noindent
where \(C\) is an arbitrary variable denoting the required velocity correction. \(\rho_{\infty_{L}}\), \(\rho_{\infty_{G}}\), \(v_{\infty_{l}}\), and \(v_{\infty_{g}}\) are taken from the freestream as it represents the edge of the boundary layer in each respective phase. Two solutions for \(C\) will result from Eq.~(\ref{eqn:TranMomBal}). The connected phase interface will tend towards the slower moving stream. Thus, the negative correction for \(C\) is neglected.

Note the deflection of the interface caused by C will decrease and the diffusion layer thickness will grow with streamwise distance. Therefore, only the closest streamwise distance considered in this paper is analyzed (i.e., \(x\) = 0.0001 m). The interface deflection can be found by taking the product of C and the time it takes for the flow to reach \(x\) = 0.0001 m, \(t\). Using the mean flow velocity, \(u_{mean}\) = 10 m/s, \(t\) = 1\(\times 10^{-5}\) s.

\begin{table}[h!]
\centering
\begin{tabular}{c c c c c}
\hline
  & p = 10 bar & p = 50 bar & p = 100 bar & p = 150 bar \\
\hline
\(|C|\) (m/s) & 2.395\(\times 10^{-2}\) & 1.488\(\times 10^{-2}\) & 1.504\(\times 10^{-2}\) & 1.595\(\times 10^{-2}\)\\
\(d_C\) (m) & 2.395\(\times 10^{-7}\) & 1.488\(\times 10^{-7}\) & 1.504\(\times 10^{-7}\) & 1.595\(\times 10^{-7}\)\\
\(\delta\) (m) & 2.311\(\times 10^{-4}\) & 6.152\(\times 10^{-5}\) & 5.152\(\times 10^{-5}\) & 4.802\(\times 10^{-5}\)\\
\hline
\end{tabular}
\caption{Velocity correction, C, interface deflection magnitude, \(d_C\), and diffusion layer thickness, \(\delta\) at \(x\) = 0.0001 m at \(p\) = 10, 50, 100, and 150 bar.}
\label{tab:fixedint}
\end{table}

Table \ref{tab:fixedint} summarizes the results. The transition between net vaporization and condensation occurs around 50 bar. At this pressure, the transverse velocity is nearly zero. As pressure deviates, the diffusion layer velocity correction increases. For all pressure cases, the interface deflection is between 2 to 3 orders of magnitude smaller than the diffusion layer thickness and can be considered negligible. Therefore, it is reasonable to assume a fixed interface at \(y\) = 0.

\section{Thermodynamic modeling and interface conditions}
\label{sec:thermo}

The governing equations need accurate estimates of the thermophysical and transport properties in a wide range of thermodynamic states to properly capture physical processes at high pressures. A real-gas equation of state is used to evaluate density, enthalpy and other thermodynamic parameters given a state point in the thermodynamic space (i.e., \(p\), \(T\) and \(Y_i\)). Transport properties are obtained from various models and correlations developed for high-pressure environments or high-dense fluids. \par

\subsection{Equation of state}
\label{subsec:eos}

In this work, the Soave-Redlich-Kwong (SRK) cubic equation of state is used~\cite{soave1972equilibrium}. From a computational perspective, a cubic equation of state is more efficient to implement than other more accurate parametric equations of state, while still providing reasonable accuracy in predicting liquid and gas solutions. However, the SRK equation of state density predictions start to deviate from experimental values as fluid density increases (e.g., liquid phase or fluids under high pressures), with errors up to 20\%~\cite{yang2000modeling,prausnitz2004thermodynamics}. To improve the accuracy of the equation of state, a volumetric correction is implemented, which recovers the molar volume, \(\bar{v}_c\), at the critical point \(T_c\) and \(p_c\) \cite{lin2006volumetric}. This method also increases the accuracy of density predictions for other temperatures and pressures. \par

The modified SRK equation of state in terms of the compressability factor, \(Z\), becomes

\begin{equation}
\label{eqn:srkmodifiedZ}
Z^3 + (3C-1)Z^2 + \big(C(3C-2)+A-B-B^2\big)Z + C(C^2-C+A-B-B^2)-AB = 0
\end{equation}

\noindent
with

\begin{equation}
Z = \frac{\bar{v}p}{R_uT} \quad ; \quad A = \frac{a(T)p}{R_{u}^{2}T^2} \quad ; \quad B = \frac{bp}{R_uT} \quad ; \quad C = \frac{c(T)p}{R_uT}
\end{equation}
\noindent
where \(R_u\) is the universal gas constant.

Eq.~(\ref{eqn:srkmodifiedZ}) is a cubic equation for \(Z\), which can be solved to obtain the molar volume or the density (\(\rho = MW/\bar{v}\)) of the mixture for a given pressure, temperature and composition. \(a(T)\) represents a temperature-dependent cohesive energy parameter, \(b\) represents the volumetric parameter and \(c(T)\) is a temperature-dependent volume correction. The cohesive parameter is evaluated from the critical point as \par


\begin{equation}
a(T) = a_c\alpha(T) \quad ; \quad a_c = \frac{1}{9(2^{1/3}-1)}\frac{R_{u}^{2}T_{c}^{2}}{p_c} 
\end{equation}

\begin{equation}
\alpha(T) = [1+m(1-T_{r}^{0.5})]^2 \quad ; \quad m = 0.48508 + 1.55171\omega-0.15613\omega^2
\end{equation}

\noindent
where \(T_r=T/T_c\) stands for the reduced temperature, \(\omega\) is the acentric factor of the species molecule, and the coefficient \(m\) is computed according to the modification proposed by Graboski and Daubert~\cite{graboski1978modified,graboski1978modified2}. The volumetric parameter and its correction are also obtained from the critical point as

\begin{equation}
b = b_c = \frac{2^{1/3}-1}{3}\frac{R_uT_c}{p_c} \quad ; \quad c(T) = c_c f(T_r) \quad ; \quad c_c = \bar{v}_{c}^{SRK}-\bar{v}_c=\bigg(\frac{1}{3}-Z_{c}^{exp}\bigg)\frac{R_uT_c}{p_c}
\end{equation}

\noindent
with \(Z_{c}^{exp}\) being the experimental compressibility factor of the critical point and \(f(T_r)\) a temperature-dependent function which becomes 1 at the critical point (\(T_r=1\)). This function is obtained from Lin et al.~\cite{lin2006volumetric}, given by

\begin{equation}
\label{eqn:Lin_model}
f(T_r) = \beta + (1-\beta)\exp\big(\eta|1-T_r|\big)
\end{equation}

In Eq.~(\ref{eqn:Lin_model}), \(\beta\) and \(\eta\) are experimentally fitted parameters for each species. To avoid isotherm crossing near the critical temperature at very high pressures, the volume correction should be temperature-independent if \(p>p_c\). Then, Eq.~(\ref{eqn:Lin_model}) is modified as in~\cite{lin2006volumetric}

\begin{equation}
\label{eqn:Lin_model2}
f(T_r) = \beta + (1-\beta)\exp\big(0.5\eta\big)
\end{equation}

For mixtures, quadratic mixing rules are used to follow the original guidelines provided by Soave~\cite{soave1972equilibrium}. Other mixing rules could be implemented, but the analysis of their performance is out of scope of the present work and satisfactory matching with experimental data has been obtained with the present model. Note that for nomenclature convenience, the dependence on temperature for the terms related to \(a\) and \(c\) is not explicitly written. The mixing rules are

\begin{equation}
\label{eqn:mixrule1}
a = \sum_{i=1}^{N}\sum_{j=1}^{N} X_i X_j(a_ia_j)^{0.5}(1-k_{ij}) \quad ; \quad b = \sum_{i=1}^{N}X_ib_i \quad ; \quad c = \sum_{i=1}^{N}X_ic_i
\end{equation}
\noindent
where \(k_i\) are experimentally fitted binary interaction parameters and \(X_i\) is the mole fraction of species \(i\). 


Evaluation of other fluid properties needed in the governing equations, such as mixture enthalpy, can be found in \ref{apn:thermo1}.






\subsection{Transport properties}
\label{subsec:transp}

The thermodynamic modeling is coupled with the computation of transport properties via high-pressure correlations that require information of the thermodynamic state of the mixture (i.e., pressure, temperature, composition and density). Viscosity and thermal conductivity are evaluated using the correlations from Chung et al.~\cite{chung1988generalized} while surface tension for the instability analysis is obtained from the Macleod-Sugden correlation, as suggested by Poling et al.~\cite{poling2001properties}. The mass diffusion coefficient, \(D_m\) is based on the model developed by Leahy-Dios and Firoozabadi~\cite{leahy2007unified}. 






The mass diffusion coefficient in Eq.~(\ref{eqn:fickian}) can be expressed as

\begin{equation}
D_m = D_{12} \Gamma_{12} \quad ; \quad \Gamma_{12} = 1 + X_1 \bigg[\bigg(\frac{\partial \ln \Phi_1}{\partial X_1}\bigg)\bigg|_{p,T}-\bigg(\frac{\partial \ln \Phi_1}{\partial X_2}\bigg)\bigg|_{p,T}\bigg]
\end{equation}

\noindent
with \(D_{12}\) computed from~\cite{leahy2007unified}. Partial derivatives of the fugacity coefficient, \(\Phi_i\), based on the SRK equation of state are shown in~\ref{apn:thermo}. Therefore, the transverse diffusion mass flux, \(J_{i}^{y}\), in Eqs.~(\ref{eqn:sc_noncons}) and~(\ref{eqn:ene_noncons}) may be substituted by \(J_{i}^{y} = - \rho D_m \frac{\partial Y_i}{\partial y}\).  \par 


The thermodynamic factor, \(\Gamma_{12}\), tends to 1 for an ideal mixture and it is identical to 1 for a pure substance. However, there is no mathematical restriction for this coefficient and it may become negative for a given composition range at a specified pressure and temperature. As other authors report~\cite{krishna2016describing}, this situation of negative or reversed diffusion is associated with phase instability of the mixture and should be avoided.  \par

\subsection{Interface matching and phase equilibrium}
\label{subsec:pheq}

The solution of the governing equations is not continuous across the interface since a jump in fluid and transport properties is present. To relate both liquid and gas phases, mass, momentum and energy conservation relations are imposed. In a frame of reference moving with the interface, which is denoted by \(\Gamma\), the mass flux, \(\dot{\omega}\), Eqs.~(\ref{eqn:massflux}) and~(\ref{eqn:spmassflux}), and energy flux, Eq.~(\ref{eqn:eneflux}), crossing the interface must be continuous (i.e., fluxes normal to the interface). This corresponds to matching the governing equations in the \(y\)-direction for the mixing layer. The transverse interface velocity, \(V_\Gamma\), is an eigenvalue of the problem that can be determined by the specific boundary conditions (i.e., \(V_{\Gamma}\) = 0 for a fixed interface). \par

\begin{equation}
\label{eqn:massflux}
\dot{\omega} = \rho_g (v_g-V_\Gamma) = \rho_l (v_l-V_\Gamma)
\end{equation}

\begin{equation}
\label{eqn:spmassflux}
\rho_g Y_{gi}(v_g-V_\Gamma) + J_{gi}^{y} = \rho_l Y_{li} (v_l-V_\Gamma) + J_{li}^{y}
\end{equation}

\begin{equation}
\label{eqn:eneflux}
\dot{\omega}(h_g-h_l) = \lambda_g\Bigg(\frac{\partial T}{\partial y}\Bigg)_g - \lambda_l\Bigg(\frac{\partial T}{\partial y}\Bigg)_l + \sum_{i=1}^{N} J_{li}^{y} h_{li} - \sum_{i=1}^{N} J_{gi}^{y} h_{gi}
\end{equation}

Rearranging Eq.~(\ref{eqn:massflux}), the normal velocity jump is obtained as

\begin{equation}
\label{eqn:veljump}
v_g-v_l = \Bigg(\frac{1}{\rho_g}-\frac{1}{\rho_l}\Bigg)\dot{\omega}
\end{equation}

The streamwise momentum equation is matched under the following conditions, which state that the tangential component of the fluid velocity at the interface should be continuous (i.e., no-slip condition), Eq.~(\ref{eqn:veltang}), as well as the tangential stress or shear stress at the interface, Eq.~(\ref{eqn:shearstress}).

\begin{equation}
\label{eqn:veltang}
u_g = u_l = U_\Gamma
\end{equation}

\begin{equation}
\label{eqn:shearstress}
\mu_g\Bigg(\frac{\partial u}{\partial y}\Bigg)_g = \mu_l\Bigg(\frac{\partial u}{\partial y}\Bigg)_l
\end{equation}

Phase-equilibrium relations provide a necessary thermodynamic closure for the interface matching. Phase equilibrium is imposed through an equality in chemical potential for each species on both sides of the interface. This condition can be expressed in terms of an equality in fugacity, \(f\), ~\cite{soave1972equilibrium,poling2001properties} as

\begin{equation}
\label{eqn:pheq1}
f_{li}(T_l,p_l,X_{li}) = f_{gi}(T_g,p_g,X_{gi})
\end{equation}

\noindent
where fugacity is a function of temperature, pressure and mixture composition. Under the constant pressure assumption, the interface pressure is continuous (i.e., \(p_l = p_g = p_\Gamma = p_{\text{ch}}\)). Eq.~(\ref{eqn:pheq1}) can be rewritten in terms of the fugacity coefficient, \(\Phi_i\), defined as

\begin{equation}
\label{eqn:pheq2}
\Phi_i=\frac{f_i}{p_i}=\frac{f_i}{pX_i}
\end{equation}

\noindent
Thus, for constant pressure across the interface, phase equilibrium is now given by

\begin{equation}
\label{eqn:pheq3}
X_{li}\Phi_{li} = X_{gi}\Phi_{gi}
\end{equation}

Furthermore, the interface presents a negligible thickness of the order of nanometers~\cite{dahms2013transition,dahms2015liquid} and diffusion layers grow fast enough on both sides of the interface~\cite{poblador2018transient}. Thus, temperature can be assumed to be the same on both sides of the interface (i.e., \(T_g=T_l = T_\Gamma\)). These assumptions simplify the solution of phase equilibrium relations and a mixture composition can readily be obtained on each side of the interface. \par

\section{Numerical method}
\label{sec:nume}

\subsection{Discretization of the governing equations}
\label{subsec:discr}

Eqs.~(\ref{eqn:gc}),~(\ref{eqn:sc_noncons}),~(\ref{eqn:xmom_noncons}), and ~(\ref{eqn:ene_noncons}) are discretized using a finite difference approach. Eq. (\ref{eqn:gc}) is discretized using an implicit approach to obtain the transverse velocity field from density variations within the domain. An explicit first-order Euler method is used to discretize Eqs.~(\ref{eqn:sc_noncons}),~(\ref{eqn:xmom_noncons}), and ~(\ref{eqn:ene_noncons}) instead of high order explicit or implicit approaches \cite{he2017sharp}. The Courant-Friedrichs-Lewy (CFL) conditions are satisfied to ensure numerical stability \cite{he2017sharp,courant1928partiellen}.

The transverse velocity and diffusion mass fluxes are computed on the cell faces, while all other properties are evaluated at the node center. A second-order central-difference scheme was used to compute those variables evaluated at both the cell face and node center. Careful CFL conditions, small transverse velocities, and an explicit scheme allow for a central-difference approach that does not introduce numerical instabilities and produces similar results as upwind schemes \cite{he2017sharp}. Gradients at the cell faces are computed using a second-order approximation. However, this scheme cannot be used at the interface as a discontinuity exists. A first-order approximation would produce inaccurate results. A one-sided second-order Taylor series expansion is instead used to correctly evaluate the gradients on both the liquid and gas sides of the interface as showed in Poblador-Ibanez and Sirignano \cite{poblador2018transient}.

\subsection{Solution algorithm}
\label{subsec:algorithm}

The governing equations and matching conditions at the interface are solved in a manner similar to the method used by Poblador-Ibanez and Sirignano \cite{poblador2018transient}. However, in the current work, the streamwise velocity field at the next streamwise position is solved first and then, is used to develop the updated transverse velocity profile.

\section{Results and discussion}
\label{sec:results}

\subsection{Diffusion layer evolution}
\label{subsec:diffdev}

\begin{figure}[h!]
\centering
\begin{subfigure}{.5\textwidth}
  \centering
  \includegraphics[width=1.0\linewidth]{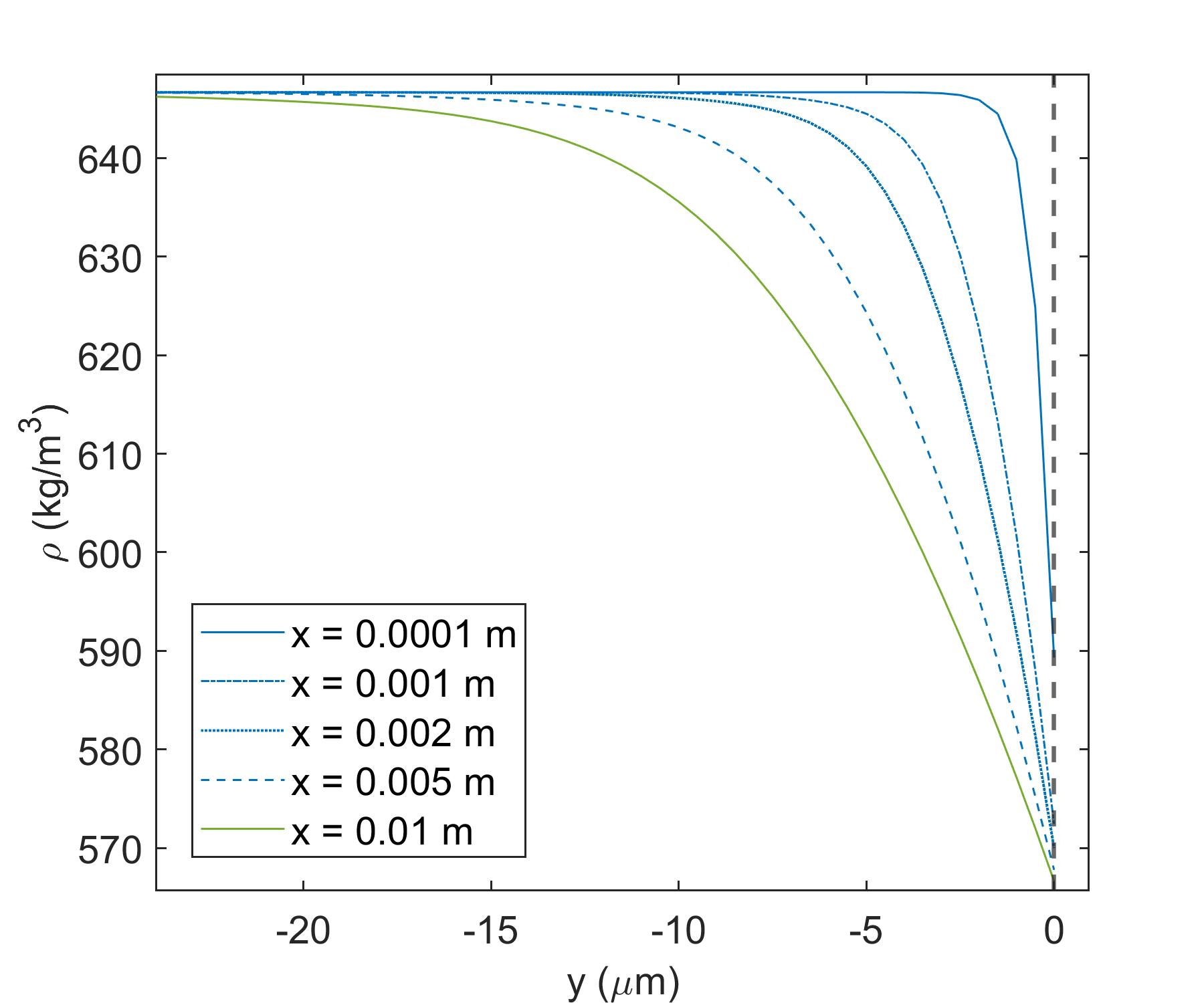}
  \caption{} 
\end{subfigure}%
\begin{subfigure}{.5\textwidth}
  \centering
  \includegraphics[width=1.0\linewidth]{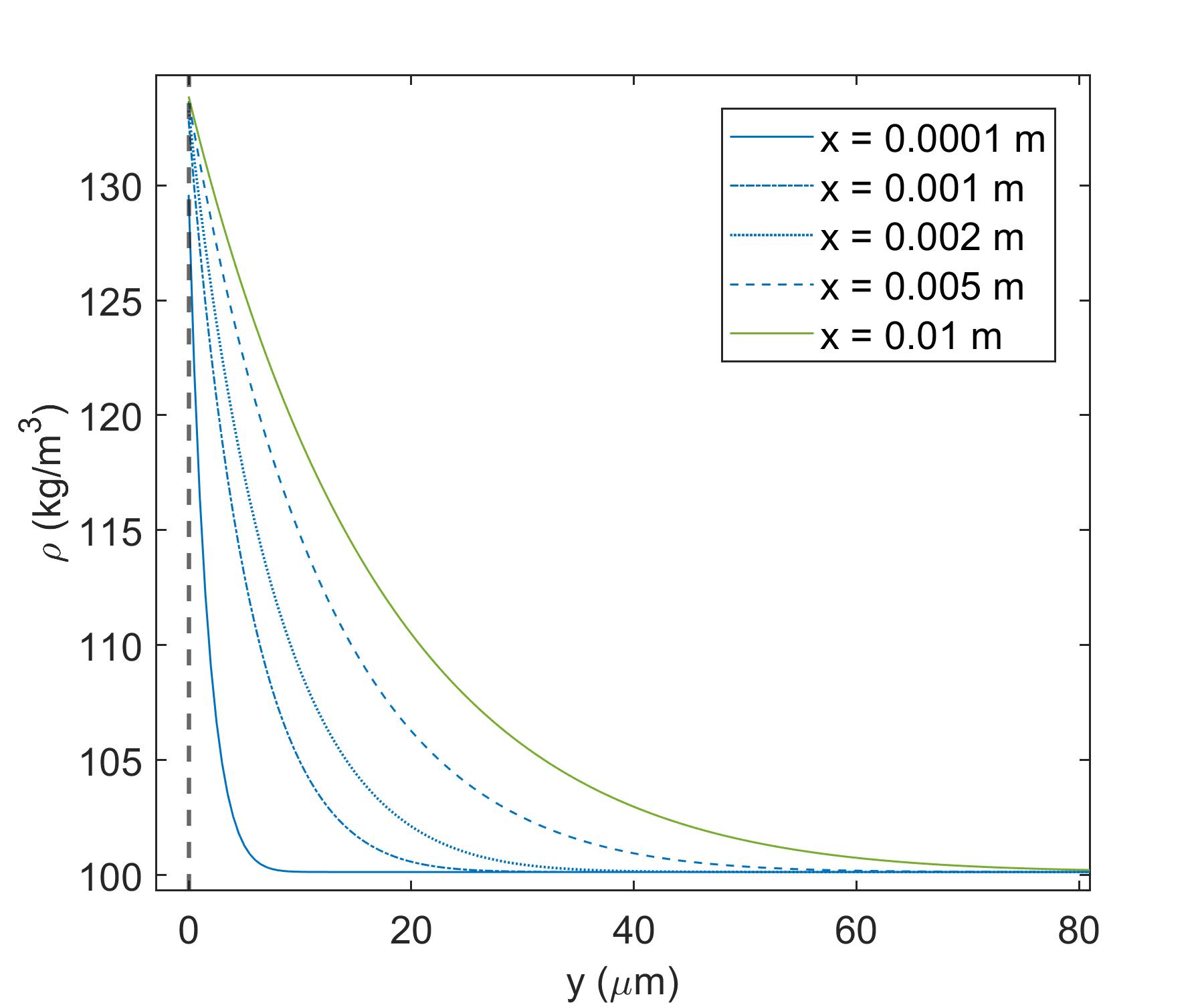}
  \caption{}
\end{subfigure}%
\
\caption{Streamwise evolution of the density distributions in the transverse direction for the oxygen/\(n\)-decane mixture at \(p =\) 150 bar. The interface is located at \(y =\) 0 \(\mu\)m. (a) liquid density; (b) gas density.}
\label{fig:den150bar}
\end{figure}

\begin{figure}[h!]
\begin{subfigure}{.5\textwidth}
  \centering
  \includegraphics[width=1.0\linewidth]{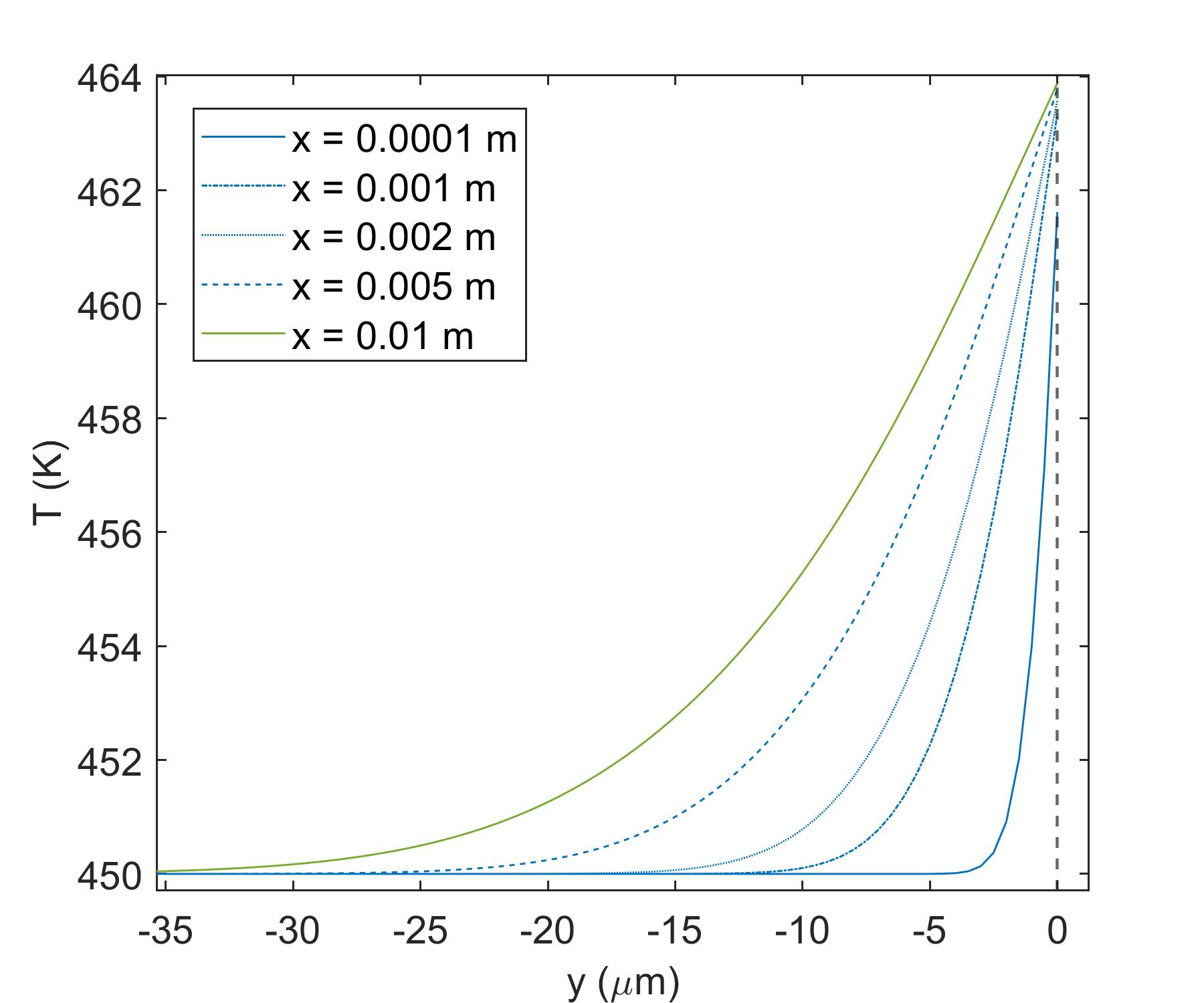}
  \caption{}
\end{subfigure}%
\begin{subfigure}{.5\textwidth}
  \centering
  \includegraphics[width=1.0\linewidth]{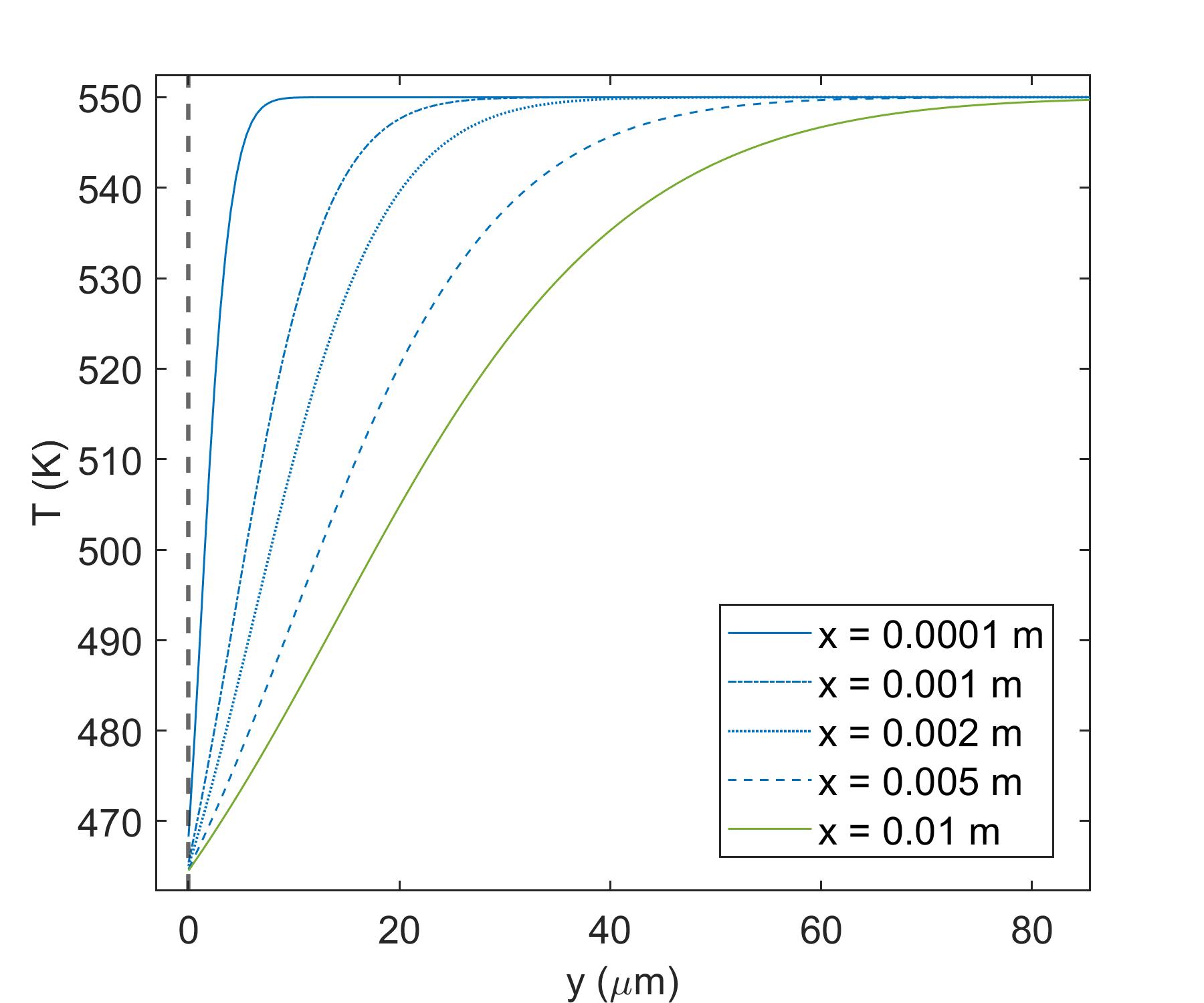}
  \caption{}
\end{subfigure}%
\caption{Streamwise evolution of the temperature distributions in the transverse direction for the oxygen/\(n\)-decane mixture at \(p =\) 150 bar. The interface is located at \(y =\) 0 \(\mu\)m. (a) liquid temperature; (b) gas temperature.}
\label{fig:temp150bar}
\end{figure}

Cold, liquid n-decane and hot, gaseous oxygen were chosen to analyze the physics within the shear layer. Figure \ref{fig:den150bar} presents the diffusion layer evolution for density at a constant pressure, \(p =\) 150 bar. At a given y-value, liquid density continuously decreases with streamwise distance as the diffusion layers grow. Conversely, the gas density increases with \(x\) at constant \(y\). This is expected because heat is transferred between the hot gas and colder liquid. In addition, heavy \(n\)-decane vaporizes into the gas phase while the lighter oxygen gas condensates into the liquid phase. As such, density will decrease in the liquid phase and increase in the gas phase as the diffusion layers grow. A comparable conclusion can be made for the temperature distributions in Figure \ref{fig:temp150bar}. The temperature in the liquid phase increases with streamwise distance whereas it decreases in the gas phase. Like the density profiles, these trends are also caused by heat conduction.


\begin{figure}[h!]
\centering
\begin{subfigure}{.5\textwidth}
  \centering
  \includegraphics[width=1.0\linewidth]{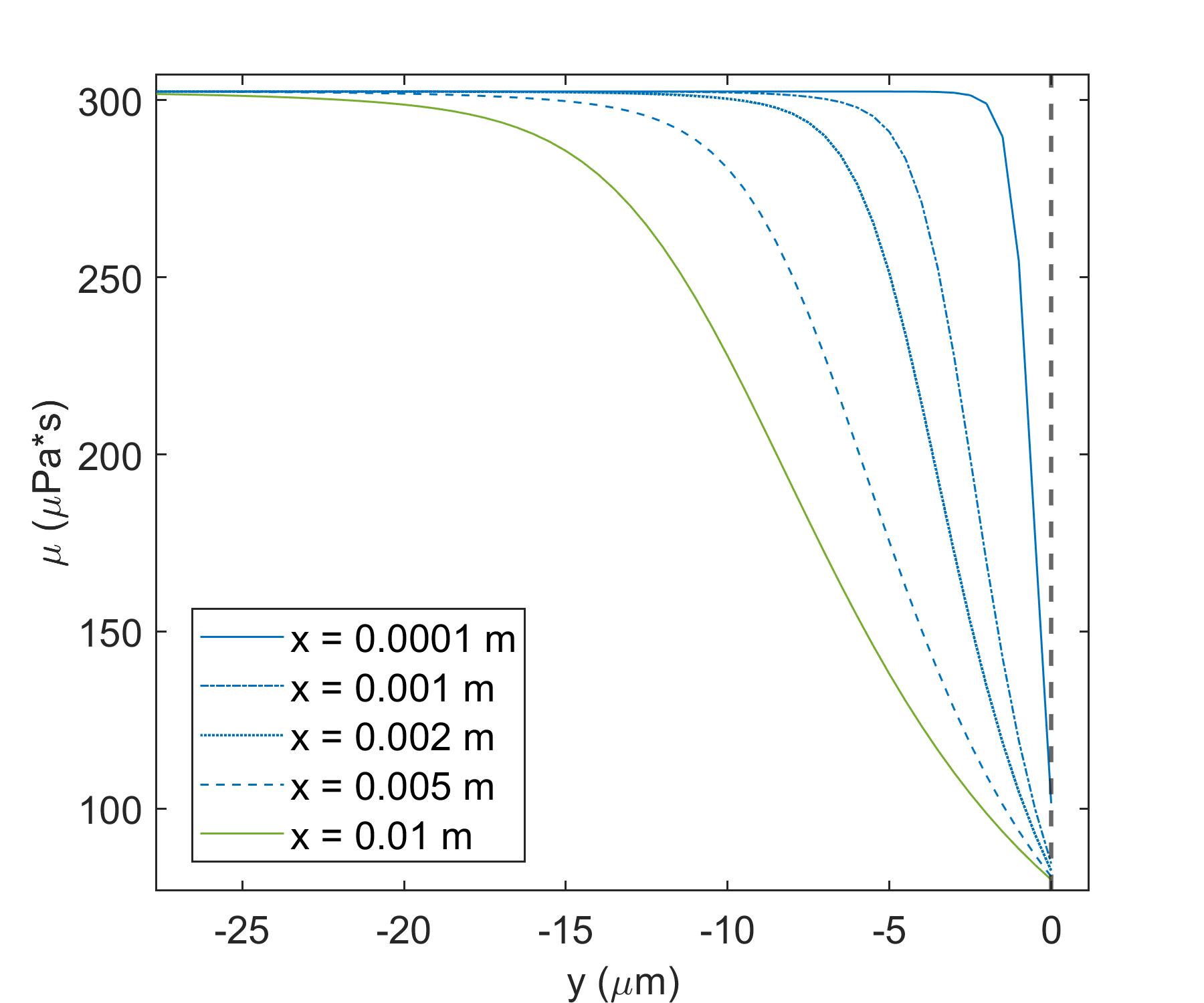}
  \caption{} 
  \label{fig:viscstream150bara}
\end{subfigure}%
\begin{subfigure}{.5\textwidth}
  \centering
  \includegraphics[width=1.0\linewidth]{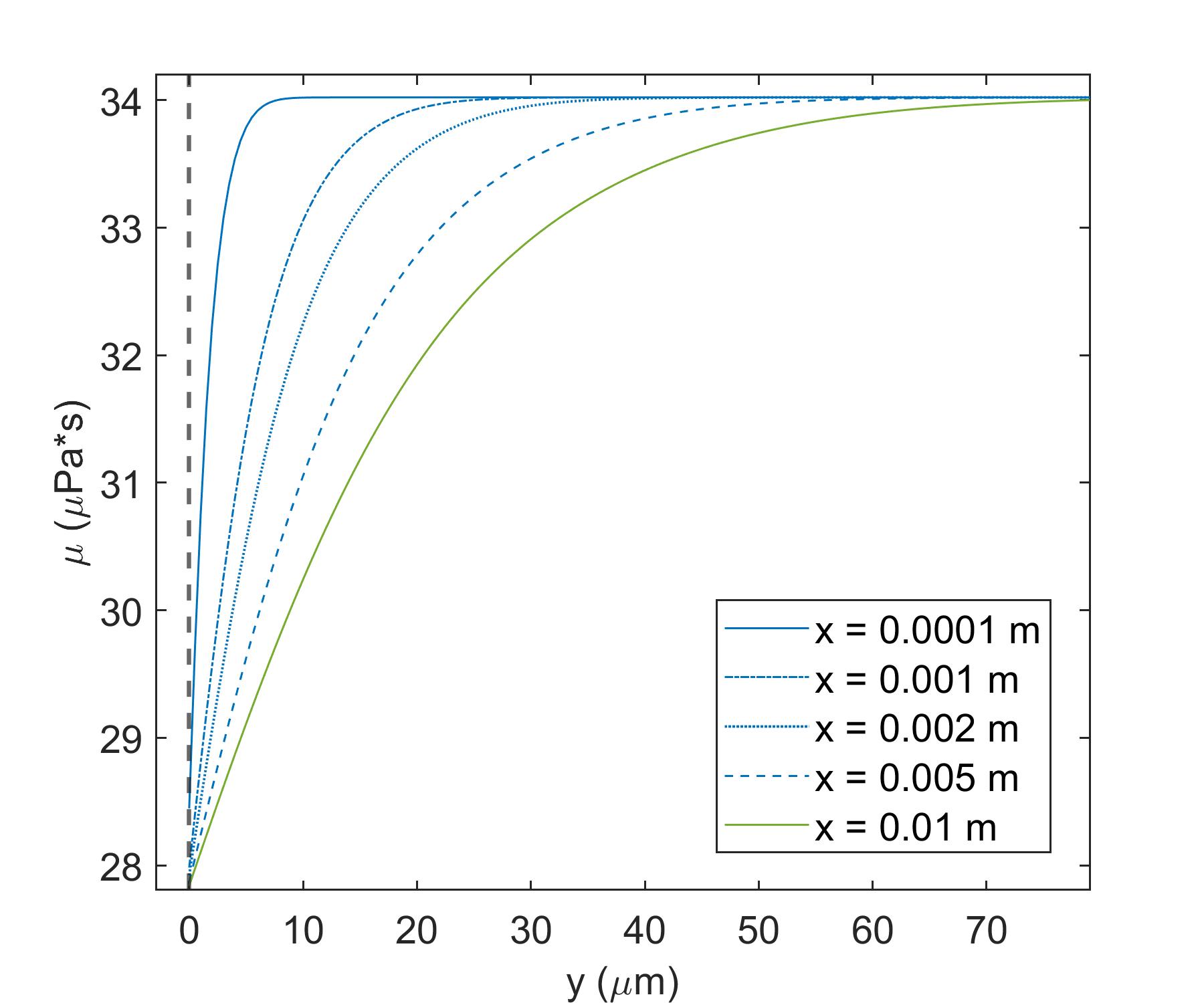}
  \caption{}
  \label{fig:viscstream150barb}
\end{subfigure}%
\caption{Streamwise evolution of the viscosity distributions in the transverse direction for the oxygen/\(n\)-decane mixture at \(p =\) 150 bar. The interface is located at \(y =\) 0 \(\mu\)m. (a) liquid viscosity; (b) gas viscosity.}
\label{fig:visc150bar}
\end{figure}

\begin{figure}[h!]
\begin{subfigure}{.5\textwidth}
  \centering
  \includegraphics[width=1.0\linewidth]{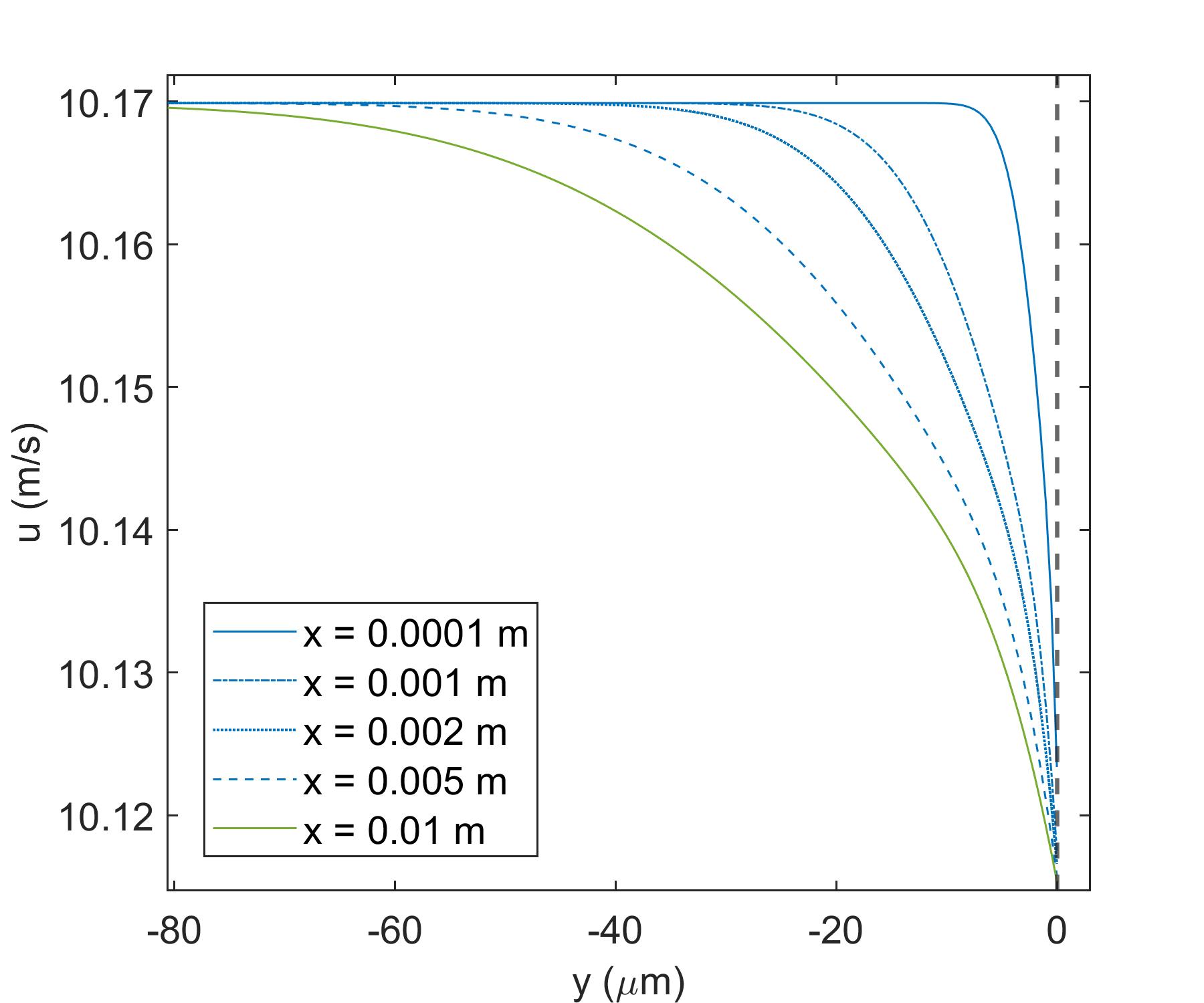}
  \caption{}
  \label{fig:viscstream150barc}
\end{subfigure}%
\begin{subfigure}{.5\textwidth}
  \centering
  \includegraphics[width=1.0\linewidth]{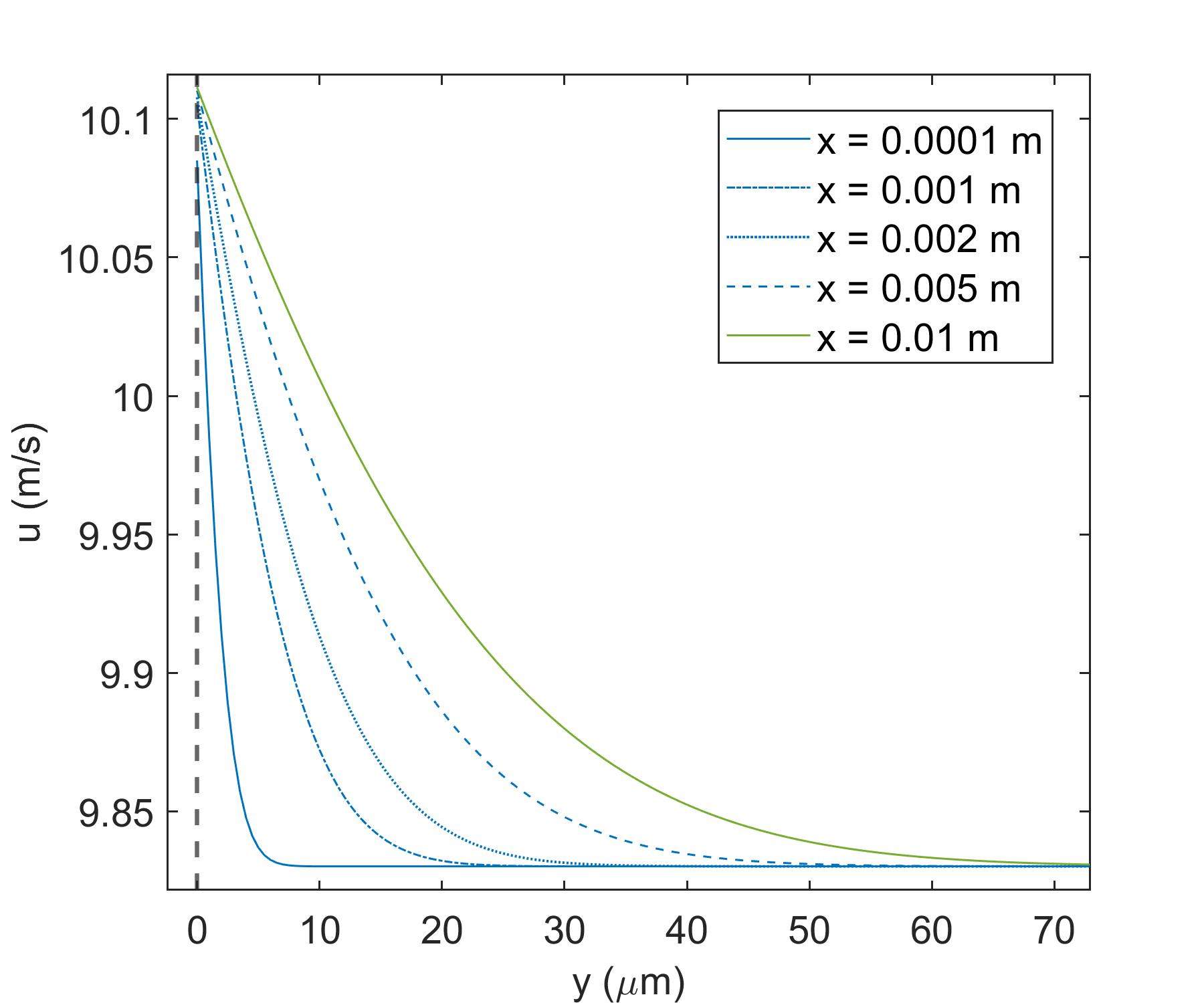}
  \caption{}
  \label{fig:viscstream150bard}
\end{subfigure}%
\caption{Streamwise evolution of the streamwise velocity distributions in the transverse direction for the oxygen/\(n\)-decane mixture at \(p =\) 150 bar. The interface is located at \(y =\) 0 \(\mu\)m. (a) liquid streamwise velocity; (b) gas streamwise velocity.}
\label{fig:stream150bar}
\end{figure}

Figure \ref{fig:visc150bar} shows the development of the viscosity profiles as the flow progresses downstream. Interestingly, the liquid-phase viscosity decreases with streamwise distance because the viscosity is dependent on temperature and density. In the gas phase, temperature has a rapid decrease as it tends towards the interface. Conversely, density increases rather minimally. Because there is a large drop in temperature across the diffusion layer in the gas phase, it overcomes density variation effects, allowing the viscosity to drop below the bulk gas viscosity. A similar result is not observed  in the liquid phase, where there is a large drop in density and an increase in temperature from the bulk conditions to the interface. Because the liquid temperature increase is slight, it has little influence on the viscosity profile.

\begin{figure}[h!]
\centering
\begin{subfigure}{.5\textwidth}
  \centering
  \includegraphics[width=1.0\linewidth]{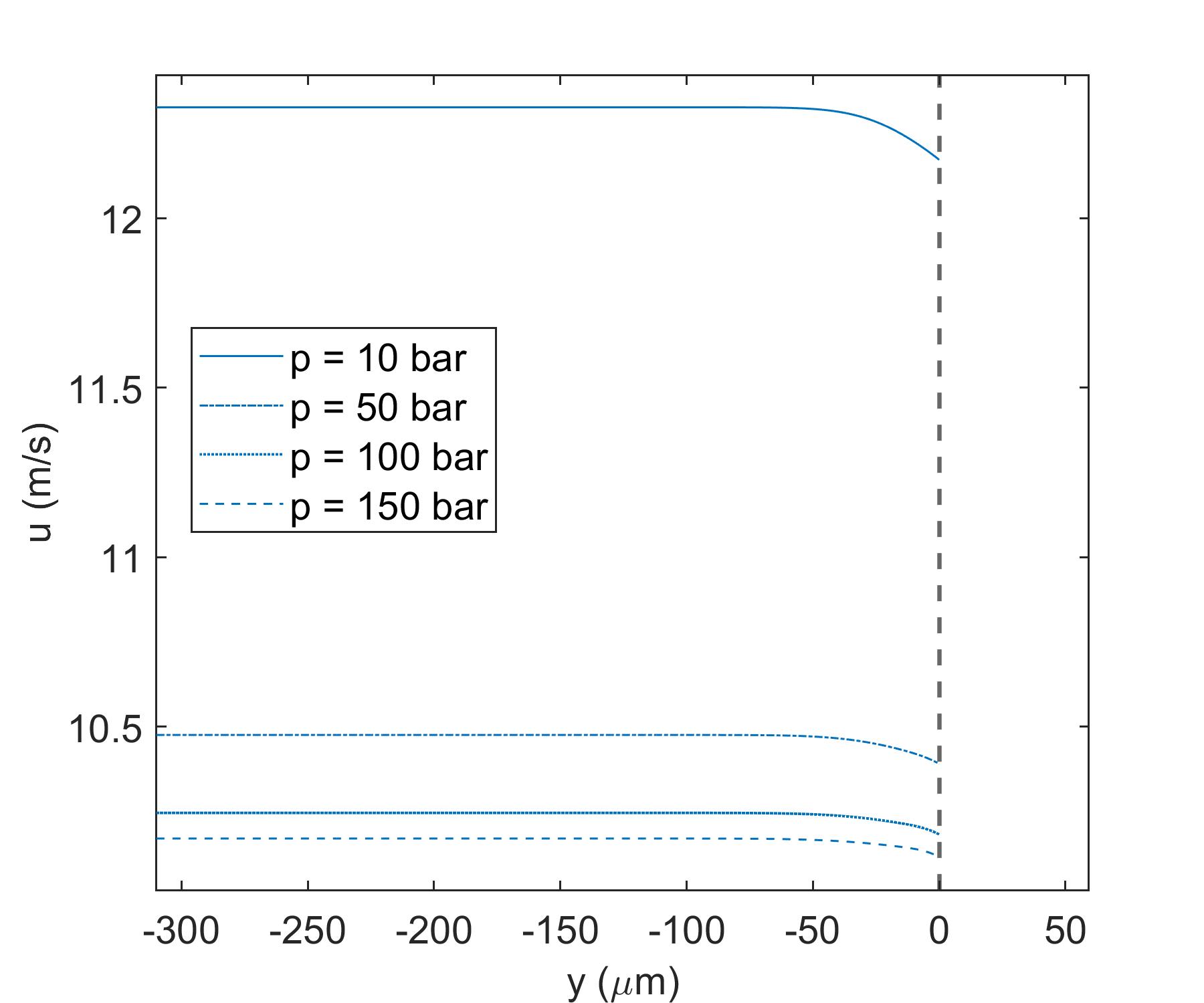}
  \caption{} 
\end{subfigure}%
\begin{subfigure}{.5\textwidth}
  \centering
  \includegraphics[width=1.0\linewidth]{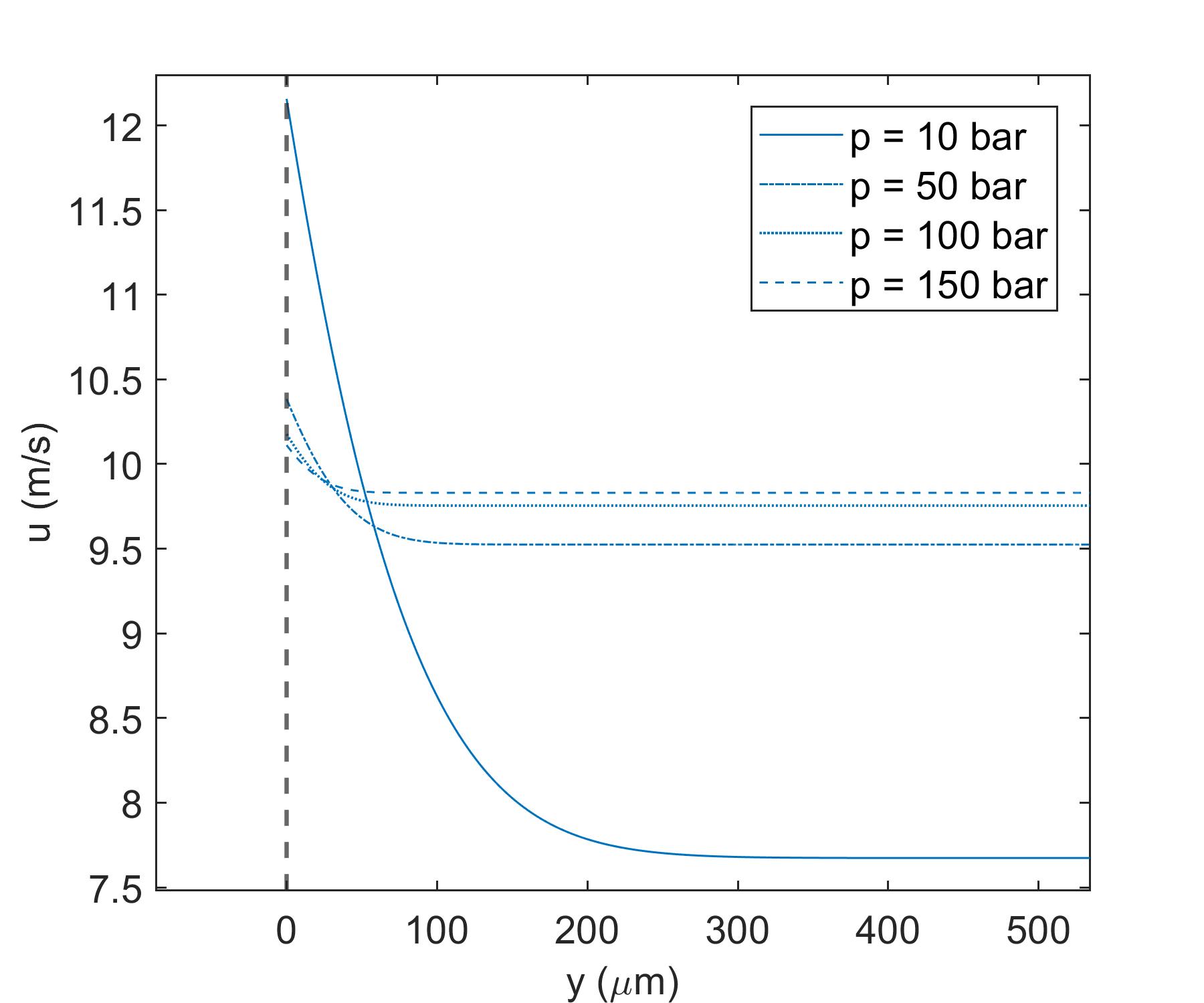}
  \caption{}
\end{subfigure}%
\
\caption{Comparison of the streamwise velocity distributions in the transverse direction for the oxygen/n-decane mixture at \(p =\) 10, 50, 100, and 150 bar and streamwise distance 1 cm (\(Re_x\) = 10,000). (a) liquid streamwise velocity; (b) gas streamwise velocity.}
\label{fig:streamVComparison}
\end{figure}

Figure \ref{fig:stream150bar} presents the streamwise velocity development. Since the bulk liquid has a larger kinematic viscosity than the bulk gas, the diffusion-layer thickness is small. The layer is much thicker on the gas side as it has a much smaller viscosity making it more readily influenced by the liquid phase. Similarly, a comparison of fully evolved streamwise velocity profiles for varying pressures can be seen from Figure \ref{fig:streamVComparison}. The layer thickness remains consistent over varying pressures in the liquid phase while the thickness progressively increases as pressure decreases in the gas phase. Density decreases as pressure decreases in both phases. Since the mean velocity of the flow remained at 10 m/s and the streamwise Reynolds number was kept constant, the velocity difference must increase for decreasing pressures. Lower densities typically correspond to larger diffusivities. Thus, momentum diffusion is larger in the low-density gas phase at low pressures. 

The interface streamwise velocity remains very close to the bulk liquid velocity. Viscosity in the gas phase changes minimally relative to the significant decrease in the liquid phase. As mentioned in the previous paragraph, the liquid is faster and more viscous than the gas. Therefore, it becomes difficult for the slower and less viscous gas to slow down the liquid stream.

\begin{table}[h!]
\centering
\begin{tabular}{c c c c c c c c c}
\hline
\(p\) (bar) & T (K) & \(X_{C_{10}H_{22},l}\) & \(X_{C_{10}H_{22},g}\) & \(\dot{\omega}\) \bigg(\(\frac{\mbox{kg}}{\mbox{m}^{2}\cdot \mbox{s}}\)\bigg) & \(\rho_l\)\bigg(\(\frac{\mbox{kg}}{\mbox{m}^{3}}\)\bigg) & \(\rho_g\) \bigg(\(\frac{\mbox{kg}}{\mbox{m}^{3}}\)\bigg) & \(h_l\) \bigg(\(\frac{\mbox{kJ}}{\mbox{kg}}\)\bigg) & \(h_g\) \bigg(\(\frac{\mbox{kJ}}{\mbox{kg}}\)\bigg)\\
\hline
10 & 451.497 & 0.974 & 0.128 & 0.0760 & 593.529 & 12.413 & 332.696 & 489.153 \\
50 & 457.610 & 0.865 & 0.0410 & 0.00488 & 580.403 & 47.932 & 355.498 & 444.624\\
100 & 461.326 & 0.744 & 0.0322 & -0.0539 & 571.833 & 91.434 & 372.589 & 437.093 \\
150 & 464.121 & 0.639 & 0.0321 & -0.109 & 563.888 & 134.362 & 386.623 & 435.013\\
\hline
\end{tabular}
\caption{Interface mean-steady temperature, mole fraction of \(n\)-decane, mass flux, density, and equilibrium mixture enthalpy in each phase for all pressure cases at streamwise distance \(x =\) 0.01 m.}
\label{tab:interfaceevo}
\end{table}

Across all pressures, the flow variables at the interface tend toward steady-state values. The effects of increasing pressure on temperature, mixture composition, and mass flux for phase change at the interface are shown in Table \ref{tab:interfaceevo}. For pressures above 50 bar, a negative mass flux is observed. This is indicative of net condensation occurring at the interface for these supercritical pressures. The dissolution of \(O_2\) in the liquid phase is enhanced by the phase-equilibrium requirements. Net vaporization occurs at the interface for the subcritical 10 bar and supercritical 50 bar pressure cases. Note that the 50 bar pressure case is very close to the transition between overall vaporization and condensation resulting in a near-zero net mass flux. Interface densities decrease with pressure in the liquid phase and increase in the gas phase. Similarly, the equilibrium mixture enthalpy increases with pressure in the liquid phase and decreases in the gas phase. Interface temperature is heavily influenced by the bulk liquid. As pressure decreases, the interface temperature's dependence on the bulk liquid temperature strengthens considerably.

\begin{figure}[h!]
\centering
\begin{subfigure}{.5\textwidth}
  \centering
  \includegraphics[width=1.0\linewidth]{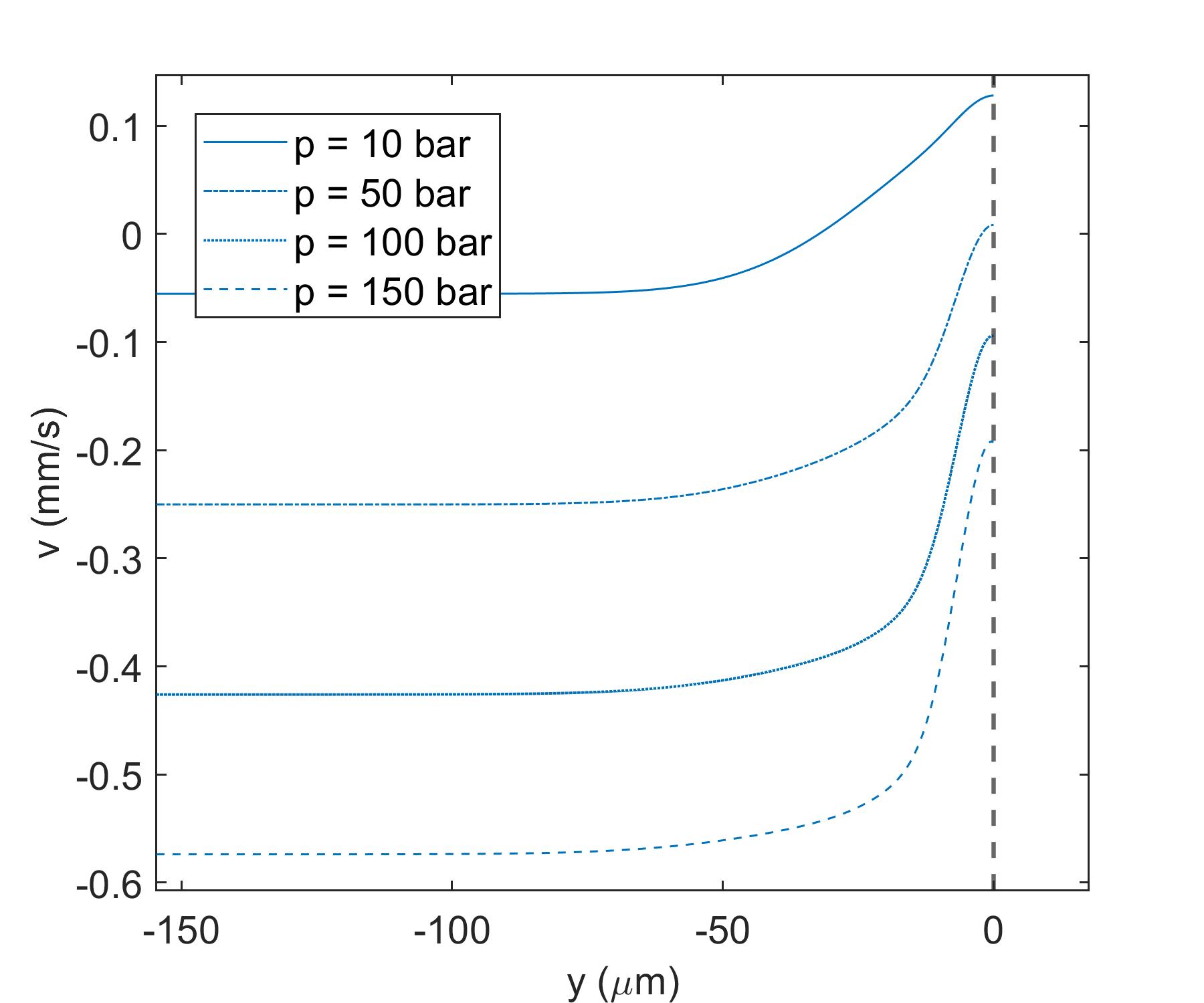}
  \caption{} 
\end{subfigure}%
\begin{subfigure}{.5\textwidth}
  \centering
  \includegraphics[width=1.0\linewidth]{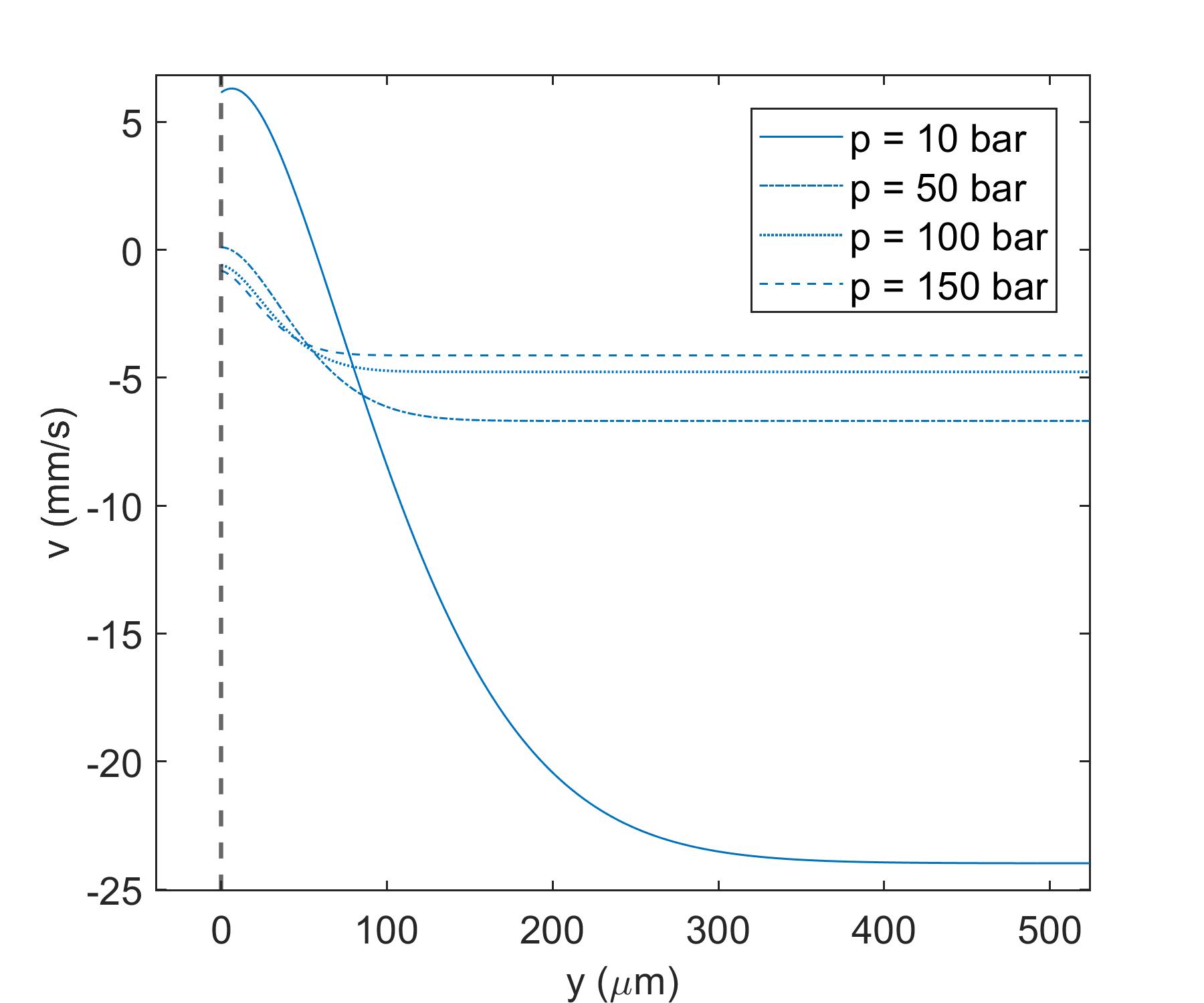}
  \caption{}
\end{subfigure}%
\caption{Comparison of the transverse velocity distributions in the transverse direction for the oxygen/n-decane mixture at \(p =\) 10, 50, 100, and 150 bar and streamwise distance 1 cm (\(Re_x\) = 10,000). (a) liquid transverse velocity; (b) gas transverse velocity.}
\label{fig:transVComparison}
\end{figure}

\begin{figure}[h!]
\begin{subfigure}{.5\textwidth}
  \centering
  \includegraphics[width=1.0\linewidth]{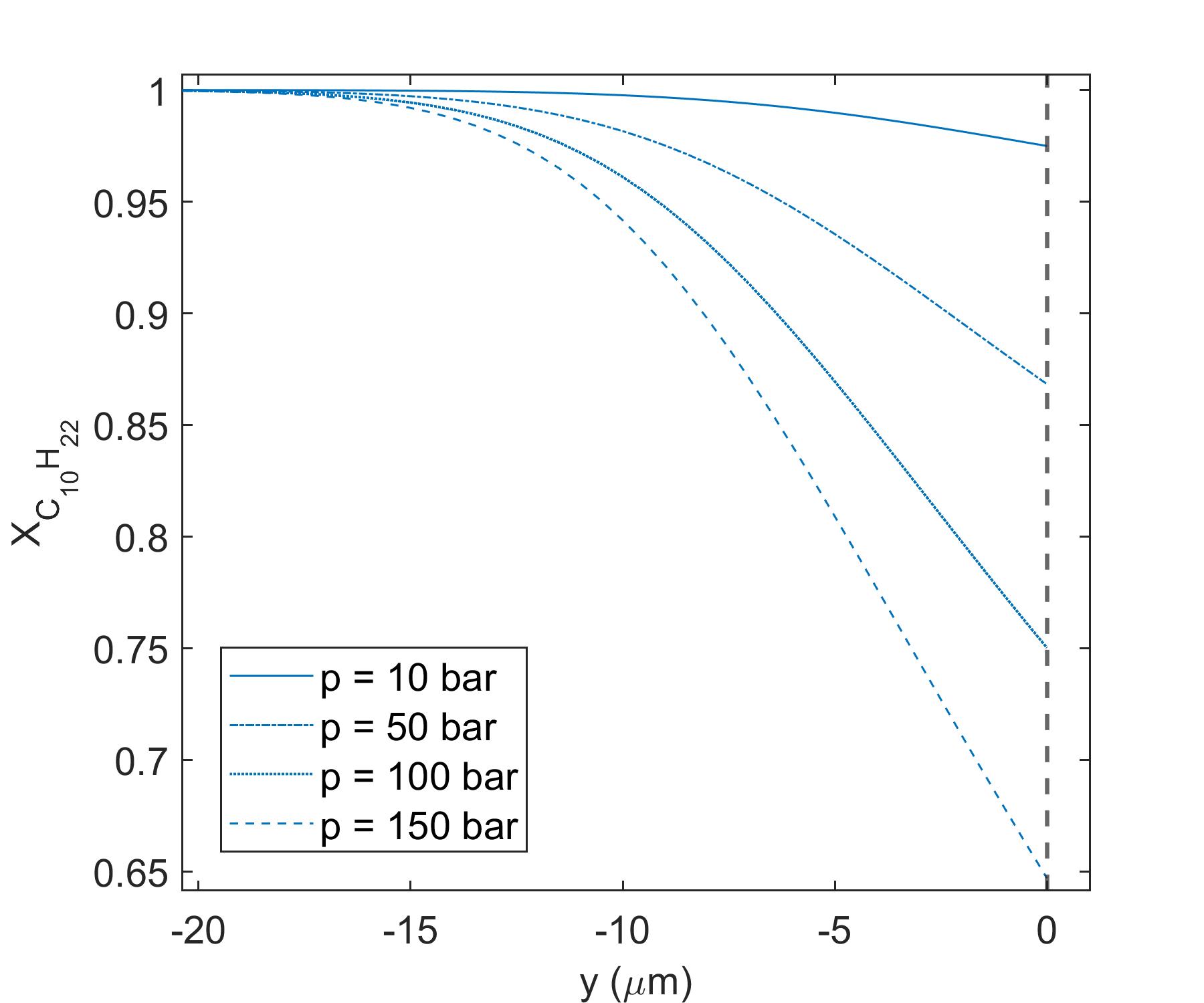}
  \caption{}
\end{subfigure}%
\begin{subfigure}{.5\textwidth}
  \centering
  \includegraphics[width=1.0\linewidth]{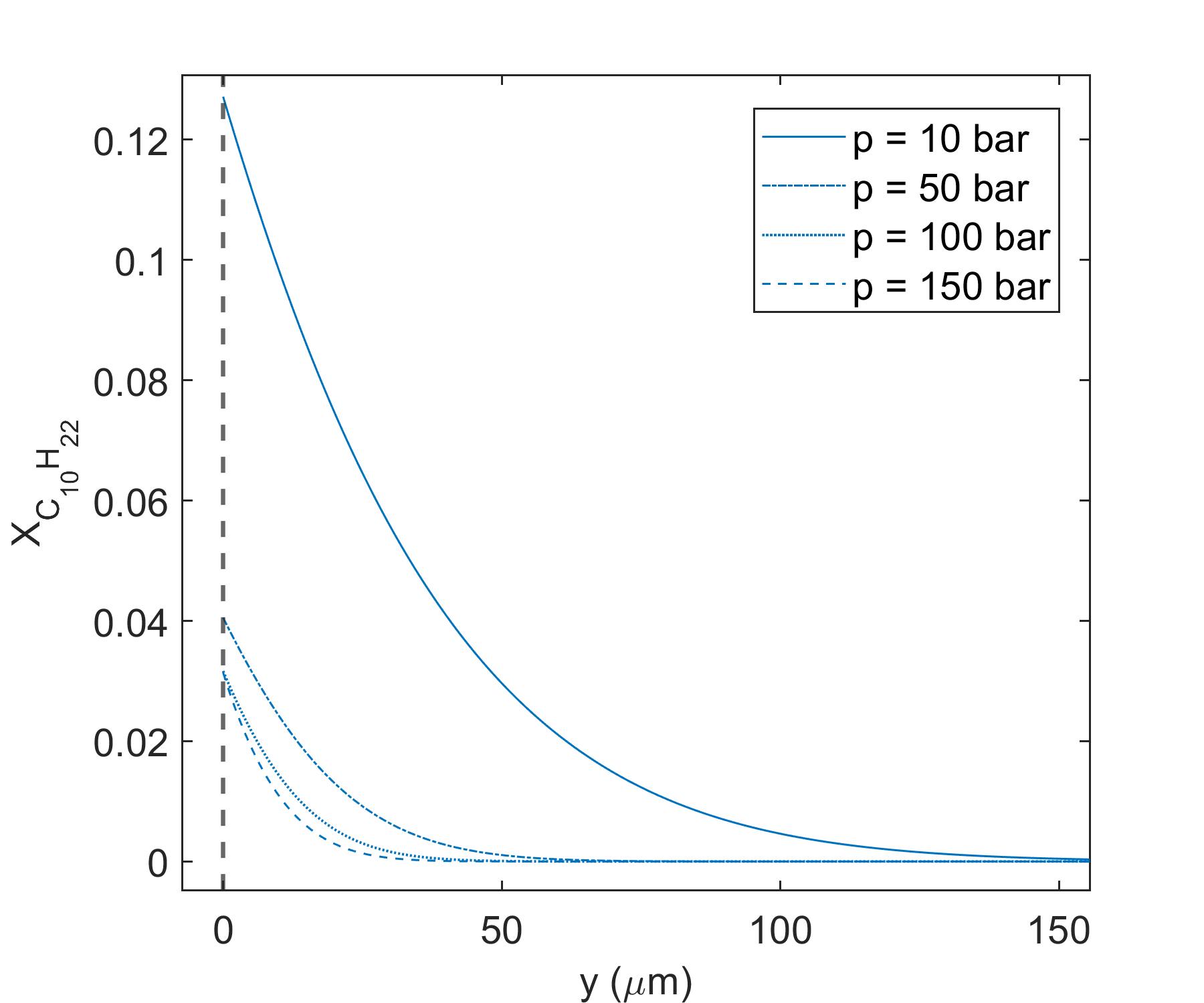}
  \caption{}
\end{subfigure}%
\caption{Comparison of the transverse velocity and mole fraction distributions in the transverse direction for the oxygen/n-decane mixture at \(p =\) 10, 50, 100, and 150 bar and streamwise distance 1 cm (\(Re_x\) = 10,000). (a) \(n\)-decane mole fraction in the liquid phase; (b) \(n\)-decane mole fraction in the gas phase.}
\label{fig:moleComparison}
\end{figure}

Figures \ref{fig:transVComparison} and \ref{fig:moleComparison} present profiles of the transverse velocity and mole fractions of \(n\)-decane at a streamwise distance \(x\) = 0.01 m for varying pressures. The diffusion layer thickness ranges from 10 - 18 \(\mu\)m in the liquid phase and 30 - 160 \(\mu\)m in the gas phase. A considerably thick diffusion layer (i.e., \(\delta =\) 160 \(\mu\)m) occurs in the gas phase at 10 bar. Similarly, the thickest diffusion layer in the liquid phase (i.e., \(\delta =\) 18 \(\mu\)m) occurs at 150 bar due to the enhanced dissolution of oxygen. The same conclusion cannot be reached for the streamwise velocity distributions in Figure \ref{fig:streamVComparison}. At 10 bar, the largest layers are witnessed on both sides of the interface. 


While Table \ref{tab:interfaceevo} and Figure \ref{fig:transVComparison} show a shift from vaporization to condensation as pressure is increased, the hotter gas still conducts heat to the colder liquid. However, the influence of energy transport by mass diffusion reverses the energy flux across the interface. That is, at high pressures either vaporization or condensation can provide the proper energy balance \cite{2017arXiv170603742Z,poblador2019analysis}.

\subsection{Similarity}
\label{subsec:simil}

Potential similarity of the solution can be seen from Figures \ref{fig:den150bar}-\ref{fig:moleComparison}. That is, reduction to one independent variable appears to be achievable. The existence of such a solution is important and useful. In such a situation, the system of partial differential equations, Eqs.~(\ref{eqn:gc})-(\ref{eqn:ene_noncons}), can be reduced to a system of ordinary differential equations that is much easier to solve. Here, a rough estimate is made concerning similarity. The approximate non-dimensional similarity variable is defined as

\begin{equation}
\label{eqn:simvar}
\eta^{*} = \frac{\sqrt{u_{\infty_{l}}}\int_{0}^{y}\rho dy^{'}}{\sqrt{2 \rho_{\infty_{l}} \mu_{\infty_{l}} x}}
\end{equation}

\(\eta^{*}\) is exact only when the product of density and viscosity is constant. Thus, here it is an approximation as \(\rho \mu\) varies across the mixing layer.

\begin{figure}[h!]
\centering
\begin{subfigure}{.5\textwidth}
  \centering
  \includegraphics[width=1.0\linewidth]{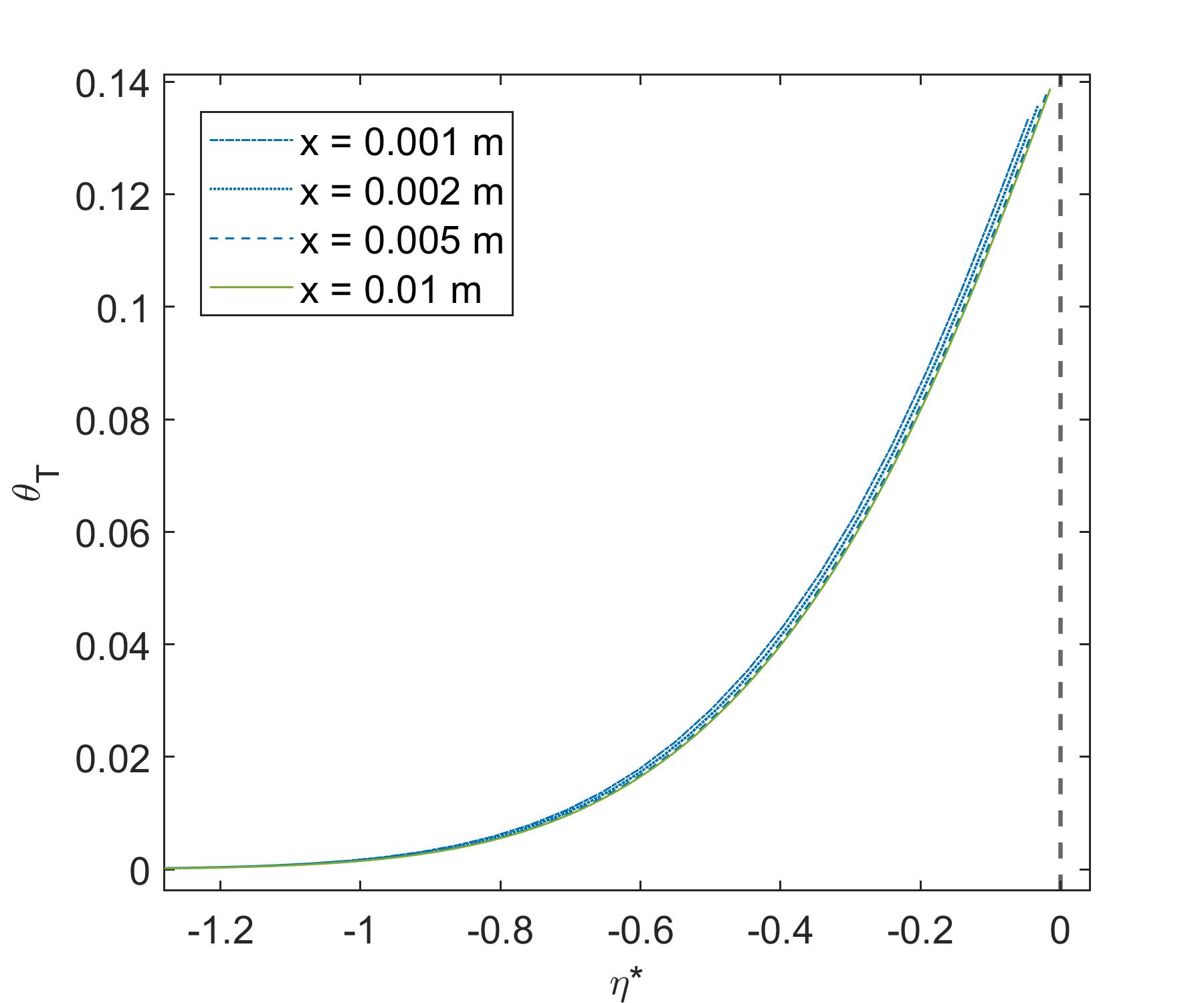}
  \caption{} 
\end{subfigure}%
\begin{subfigure}{.5\textwidth}
  \centering
  \includegraphics[width=1.0\linewidth]{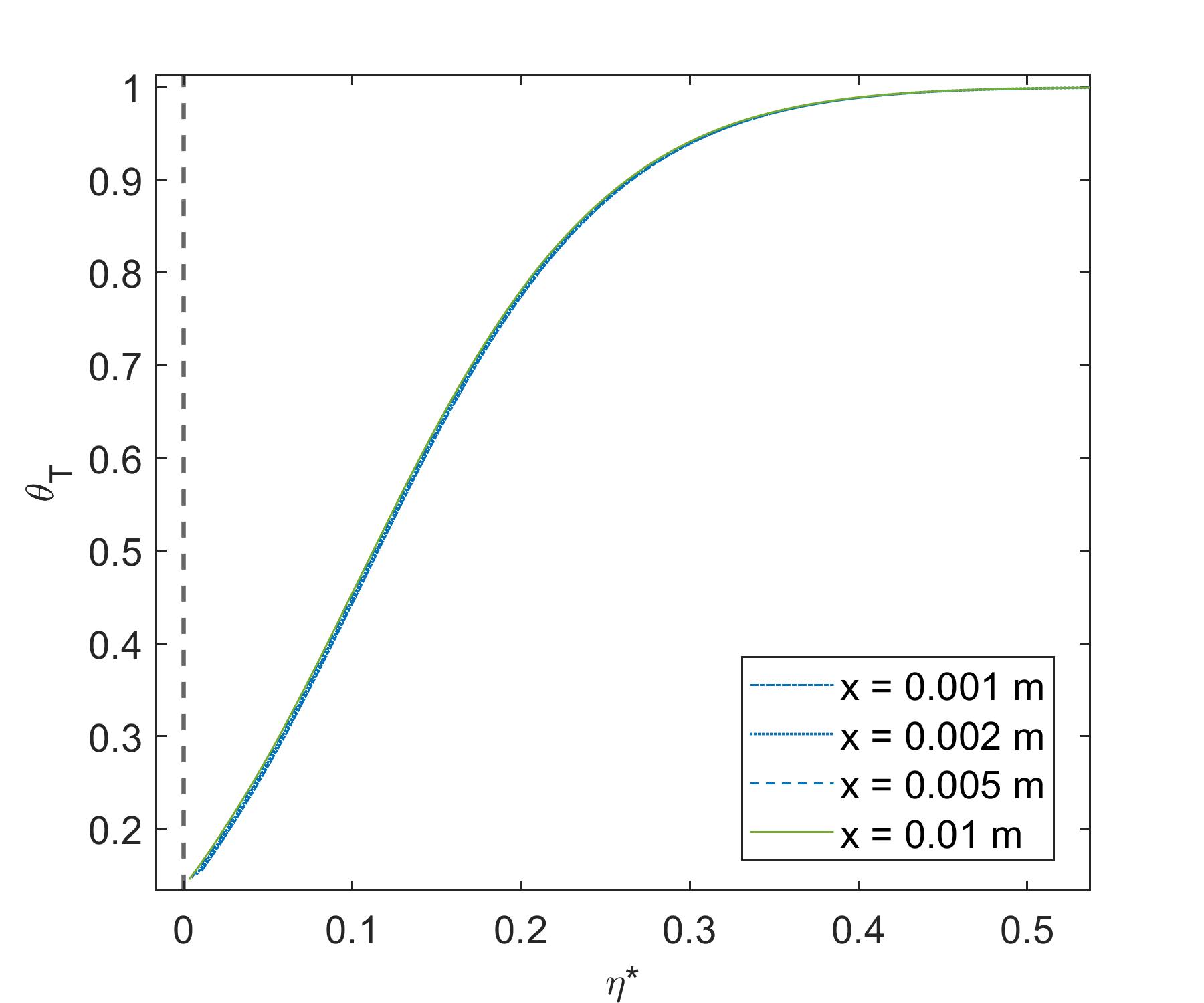}
  \caption{}
\end{subfigure}%
\caption{Evolution of the non-dimensional temperature distributions in the transverse direction against the non-dimensional similarity variable, \(\eta^*\), for the oxygen/n-decane mixture at \(p =\) 150 bar. (a) liquid temperature; (b) gas temperature.}
\label{fig:tempsim150}
\end{figure}

\begin{figure}[h!]
\begin{subfigure}{.5\textwidth}
  \centering
  \includegraphics[width=1.0\linewidth]{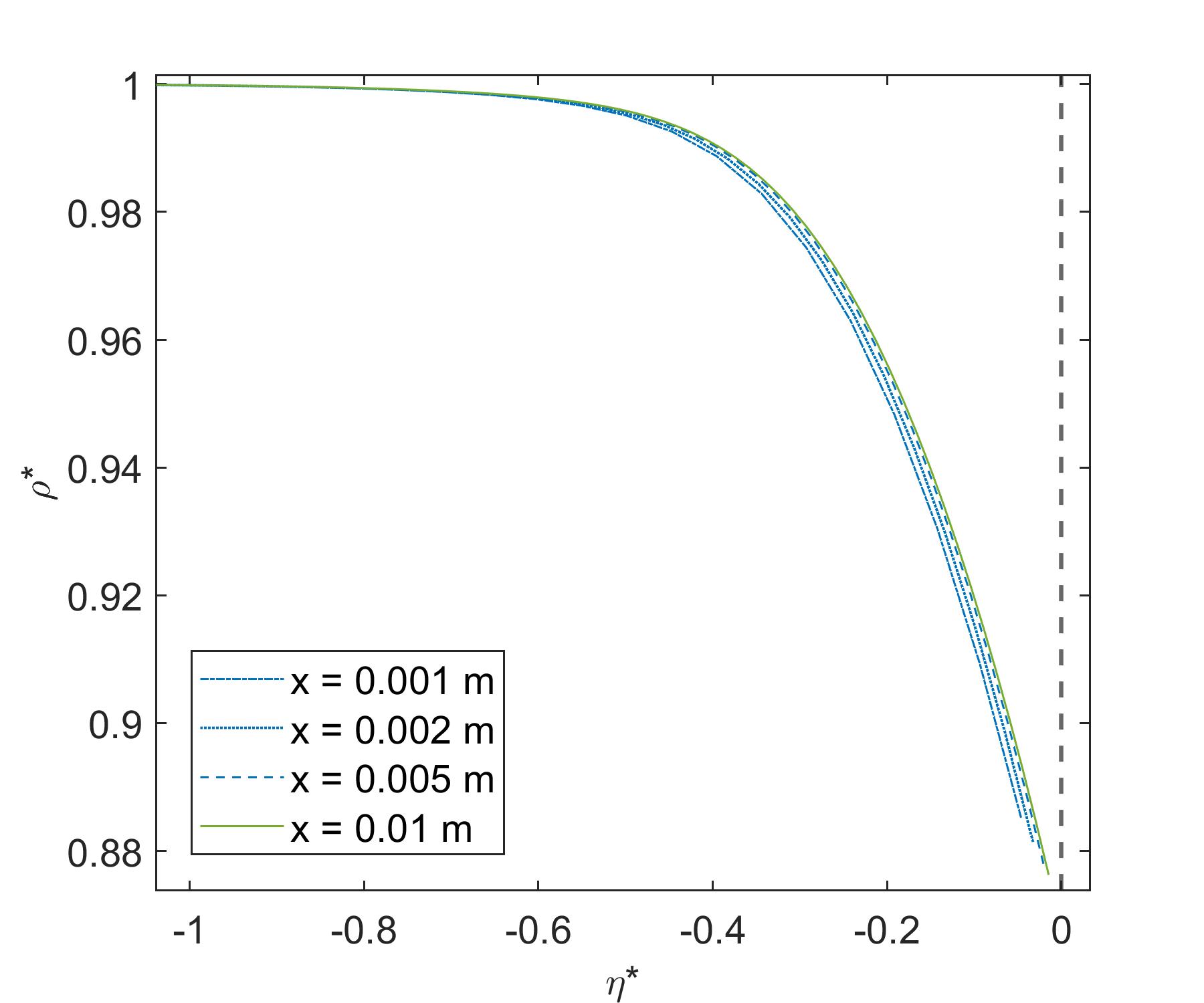}
  \caption{}
\end{subfigure}%
\begin{subfigure}{.5\textwidth}
  \centering
  \includegraphics[width=1.0\linewidth]{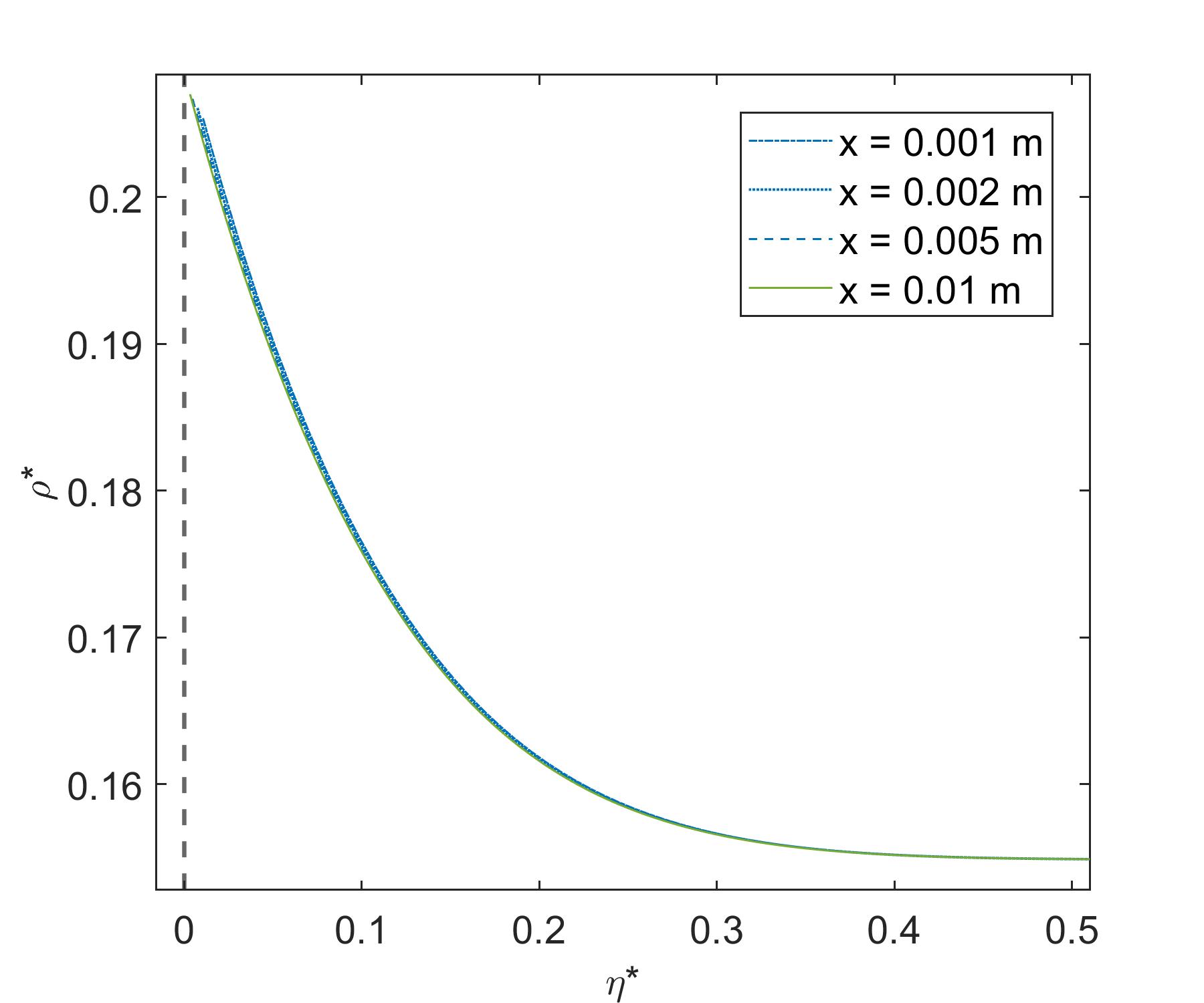}
  \caption{}
\end{subfigure}%
\caption{Evolution of the non-dimensional temperature and density distributions in the transverse direction against the non-dimensional similarity variable, \(\eta^*\), for the oxygen/n-decane mixture at \(p =\) 150 bar. (a) liquid density; (b) gas density.}
\label{fig:densim150}
\end{figure}

\begin{figure}[h!]
\centering
\begin{subfigure}{.5\textwidth}
  \centering
  \includegraphics[width=1.0\linewidth]{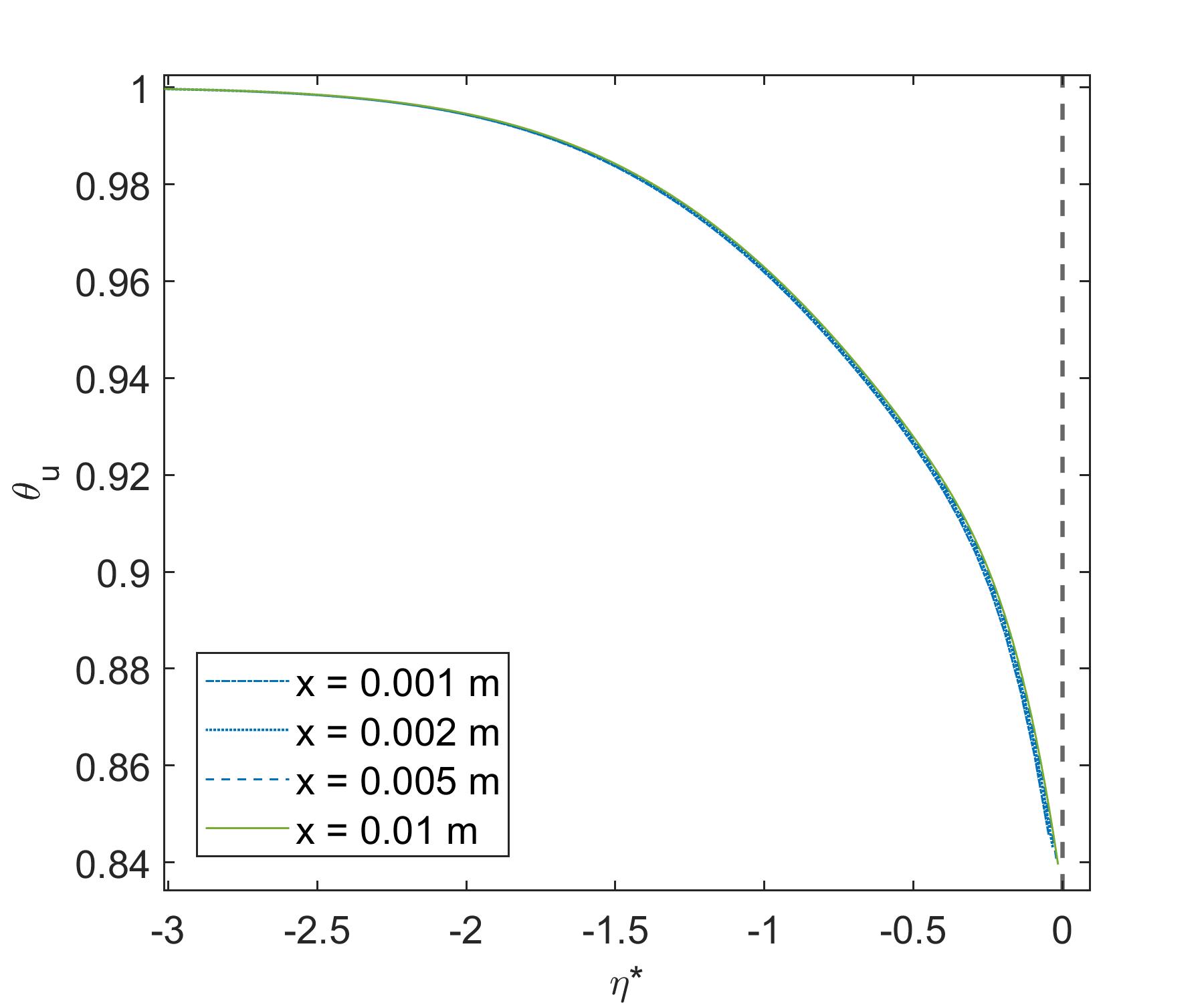}
  \caption{} 
\end{subfigure}%
\begin{subfigure}{.5\textwidth}
  \centering
  \includegraphics[width=1.0\linewidth]{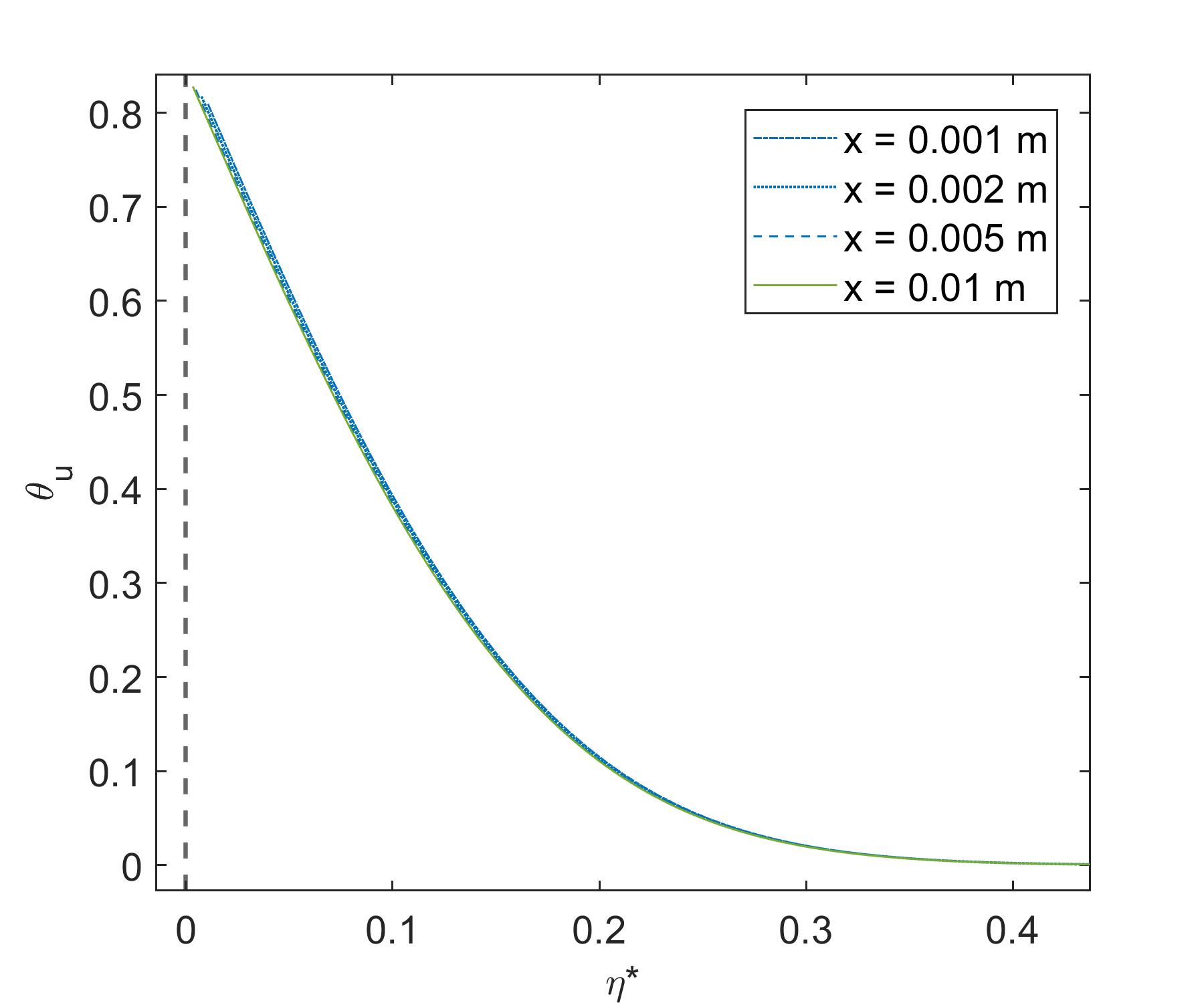}
  \caption{}
\end{subfigure}%
\
\caption{Evolution of the non-dimensional streamwise velocity distribution in the transverse direction against the non-dimensional similarity variable, \(\eta^*\), for the oxygen/n-decane mixture at \(p =\) 150 bar. (a) streamwise velocity in the liquid phase near the interface; (b) streamwise velocity in the gas phase near the interface.}
\label{fig:streamtranssim150}
\end{figure}

Figures \ref{fig:tempsim150}-\ref{fig:streamtranssim150} present plots of the similarity profile development for temperature, density, and streamwise velocity at \(p =\) 150 bar. Continuous variables, such as streamwise velocity and temperature, are non-dimensionalized to obtain distributions ranging from 0 to 1 as

\begin{equation}
\theta_{u}(y) = \frac{u(y)-u_{\infty_{g}}}{u_{\infty_{l}}-u_{\infty_{g}}} \quad ; \quad \theta_{T}(y) = \frac{T(y)-T_{\infty_{l}}}{T_{\infty_{g}}-T_{\infty_{l}}}
\label{eqn:normalizedContinuous}
\end{equation}
Similarly, density is non-dimensionalized with respect to the liquid freestream conditions as

\begin{equation}
\rho^{*}= \frac{\rho}{\rho_{\infty_{l}}}
\label{eqn:normalizedDiscontinuous}
\end{equation}

\noindent
For all flow variables, the diffusing-advecting quantities collapse to a near-similar solution. 

\begin{figure}[h!]
\centering
\begin{subfigure}{.5\textwidth}
  \centering
  \includegraphics[width=1.0\linewidth]{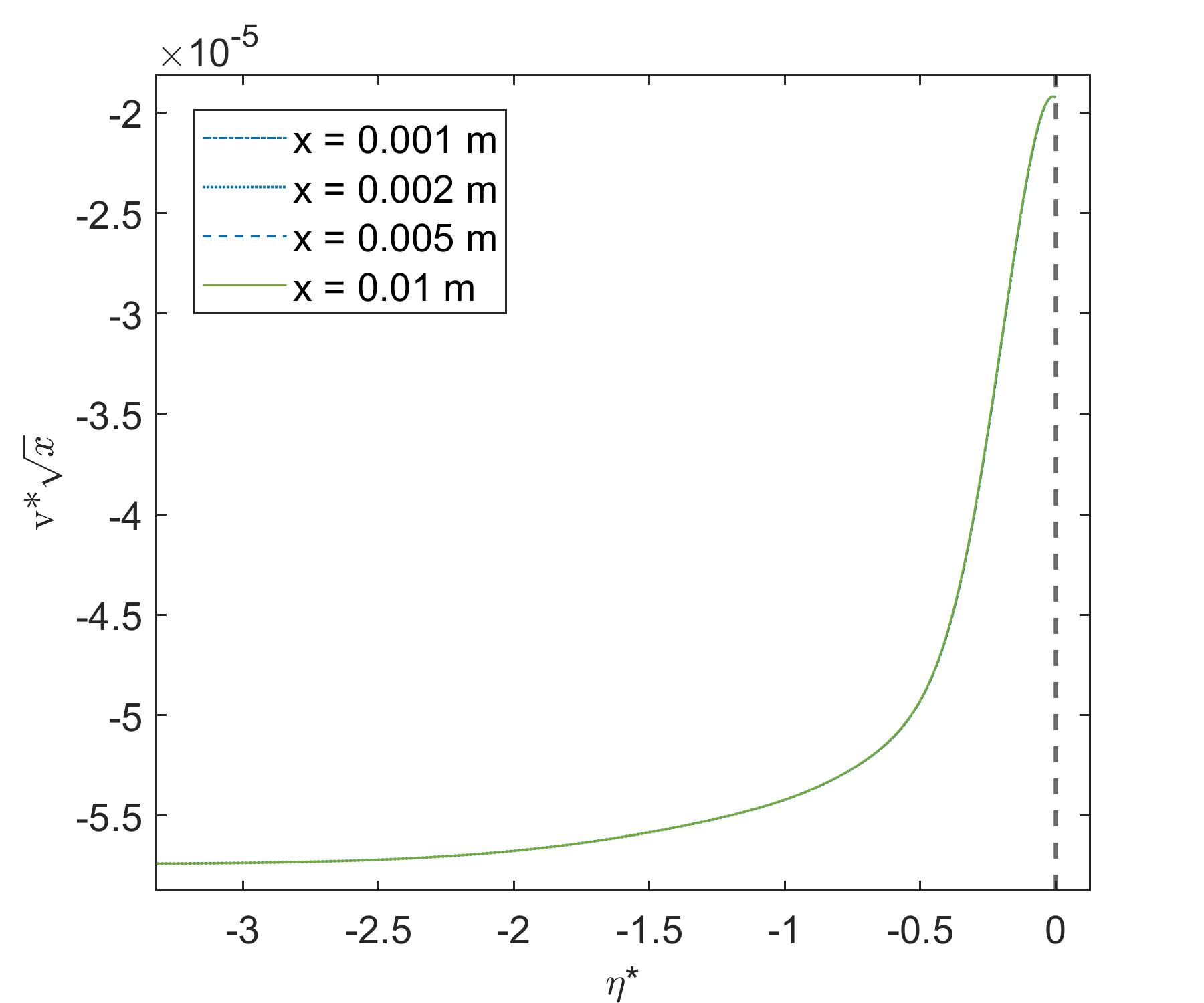}
  \caption{} 
\end{subfigure}%
\begin{subfigure}{.5\textwidth}
  \centering
  \includegraphics[width=1.0\linewidth]{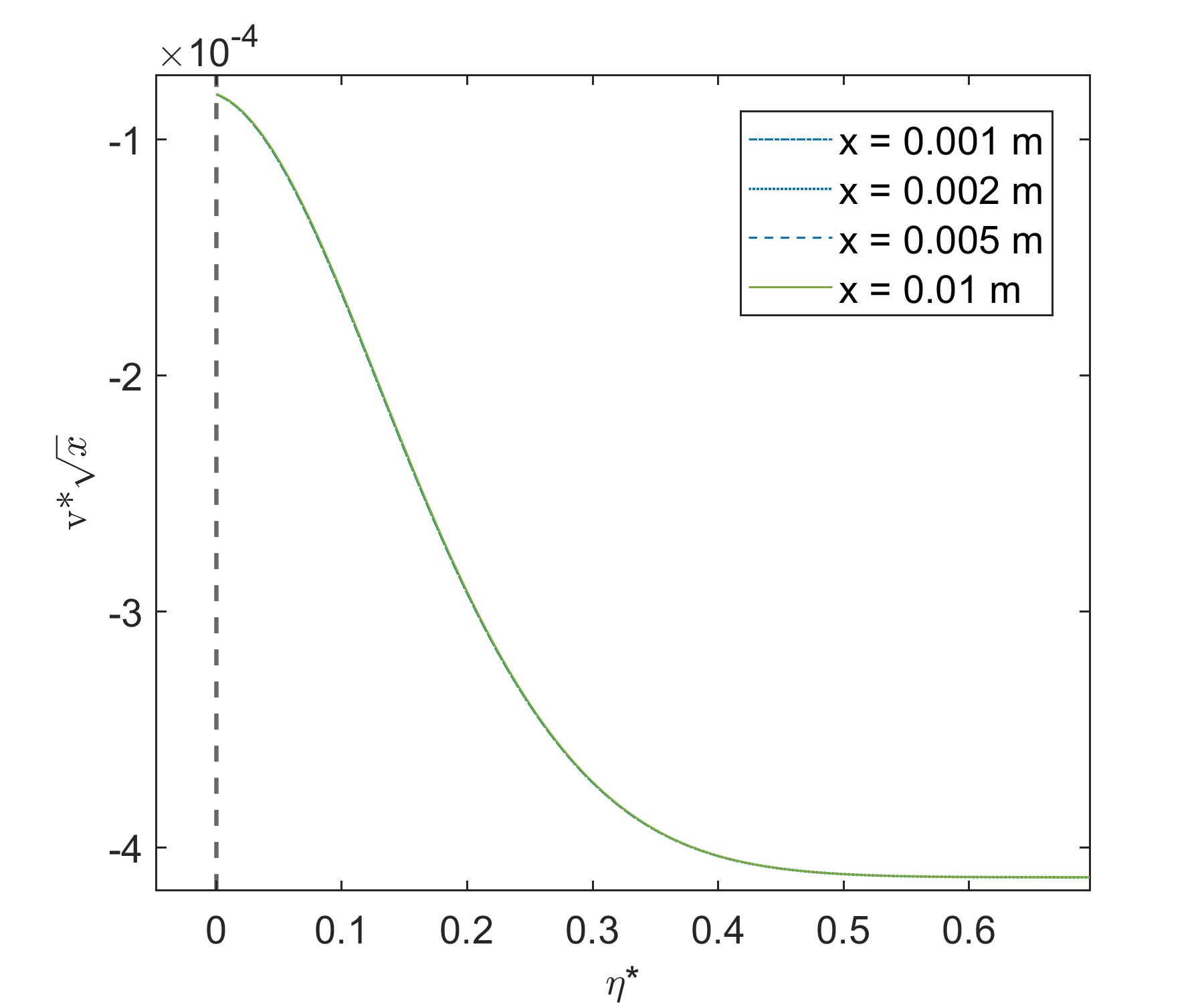}
  \caption{}
\end{subfigure}%
\
\caption{Evolution of the streamwise distance-weighted transverse velocity distribution in the transverse direction against the non-dimensional similarity variable, \(\eta^*\), for the oxygen/n-decane mixture at \(p =\) 150 bar. (a) weighted transverse velocity in the liquid phase near the interface; (b) weighted transverse velocity in the gas phase near the interface.}
\label{fig:transsim150}
\end{figure}

Similarity for the transverse velocity can be analyzed by multiplying the transverse velocity profiles by the square root of their respective streamwise distances. That is, the profile dependence on streamwise distance \(x\) is removed. Figure \ref{fig:transsim150} presents the weighted transverse velocity against \(\eta*\). Like the other variable distributions, a similar solution is reached for all streamwise locations. 


\begin{figure}[h!]
\centering
\begin{subfigure}{.5\textwidth}
  \centering
  \includegraphics[width=1.0\linewidth]{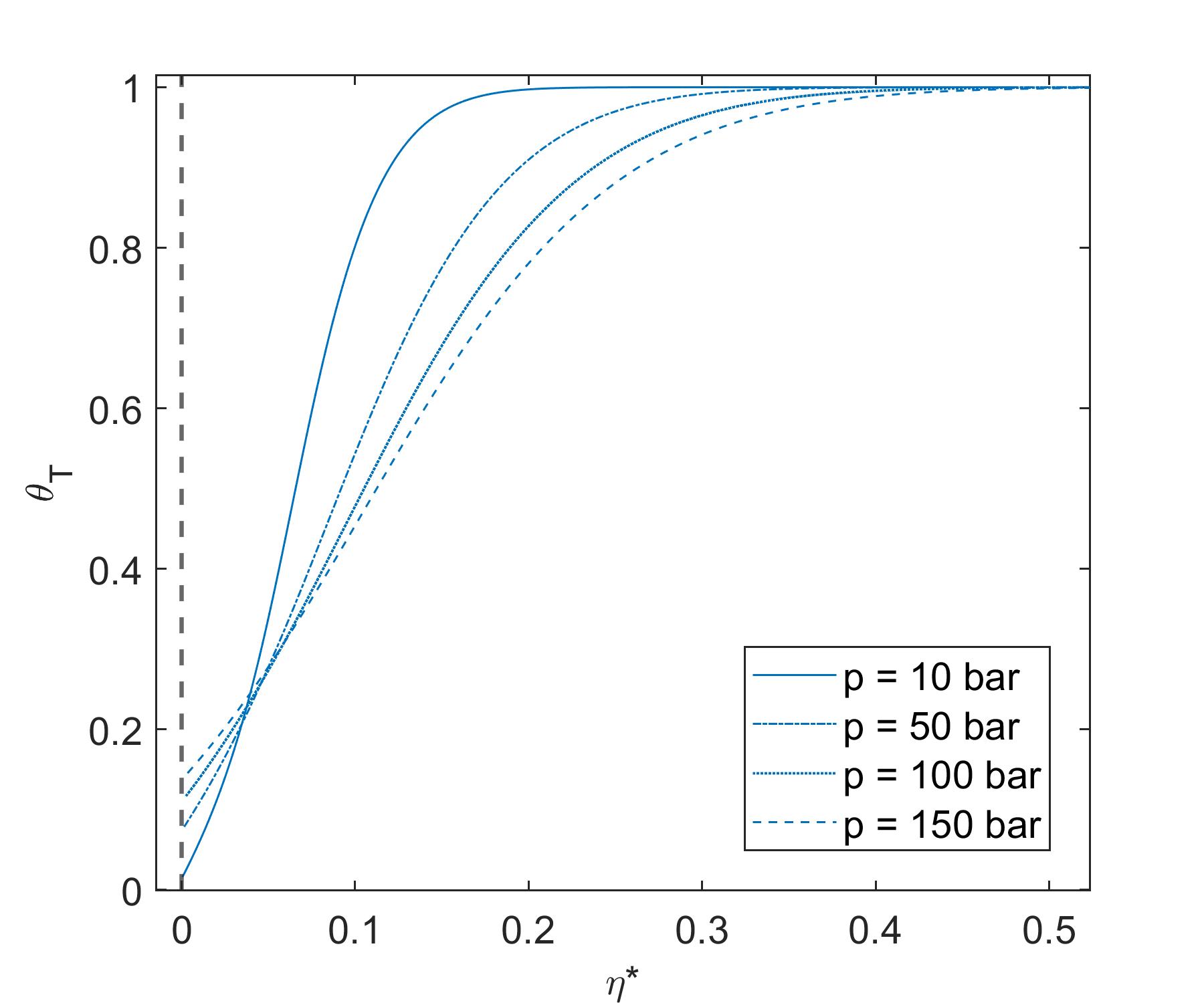}
  \caption{} 
\end{subfigure}%
\begin{subfigure}{.5\textwidth}
  \centering
  \includegraphics[width=1.0\linewidth]{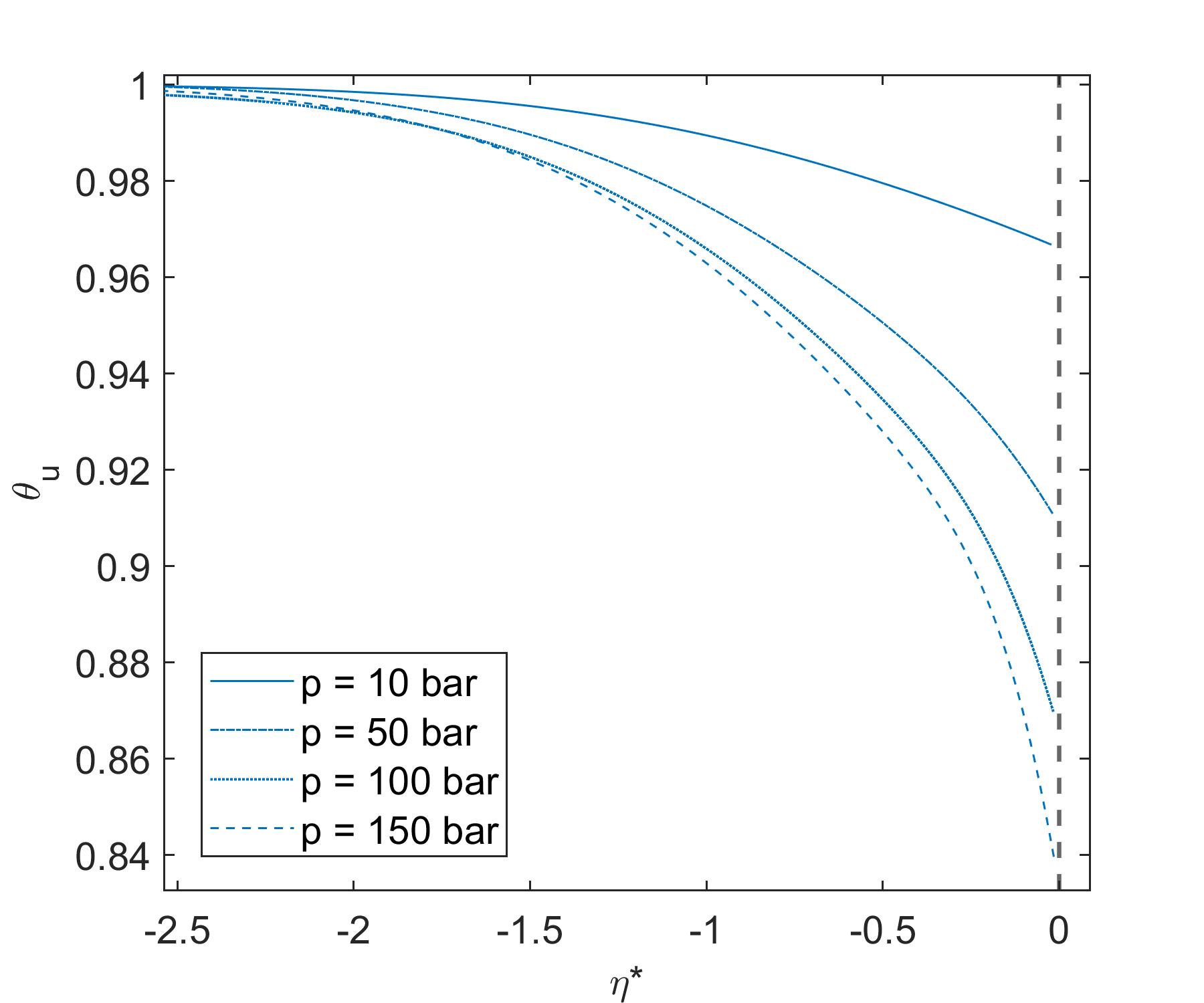}
  \caption{}
\end{subfigure}%
\
\caption{Comparison of the non-dimensional temperature and streamwise velocity distributions in the transverse direction against the similarity variable, \(\eta^*\), for the oxygen/n-decane mixture at \(p =\) 10, 50, 100, and 150 bar and streamwise position \(x = 1\) cm. (a) normalized temperature in the gas phase near the interface; (b) normalized streamwise velocity in the liquid phase near the interface.}
\label{fig:tempstreamLVarP}
\end{figure}

Across different pressure cases, the similarity profiles differ. Figure \ref{fig:tempstreamLVarP} shows the considerable deviation of the profiles of non-dimensionalized temperature and streamwise velocity for different pressures. This is also the case for the streamwise velocity. The interface temperature and velocity are significantly changed because of the altered thermodynamics. In addition, the diffusion layer evolution and thickness are dependent on the density, viscosity, and streamwise velocity difference between the two phases. All of these parameters change as pressure changes and hence, the streamwise velocity profiles also vary.  

\begin{figure}[h!]
\centering
\begin{subfigure}{.5\textwidth}
  \centering
  \includegraphics[width=1.0\linewidth]{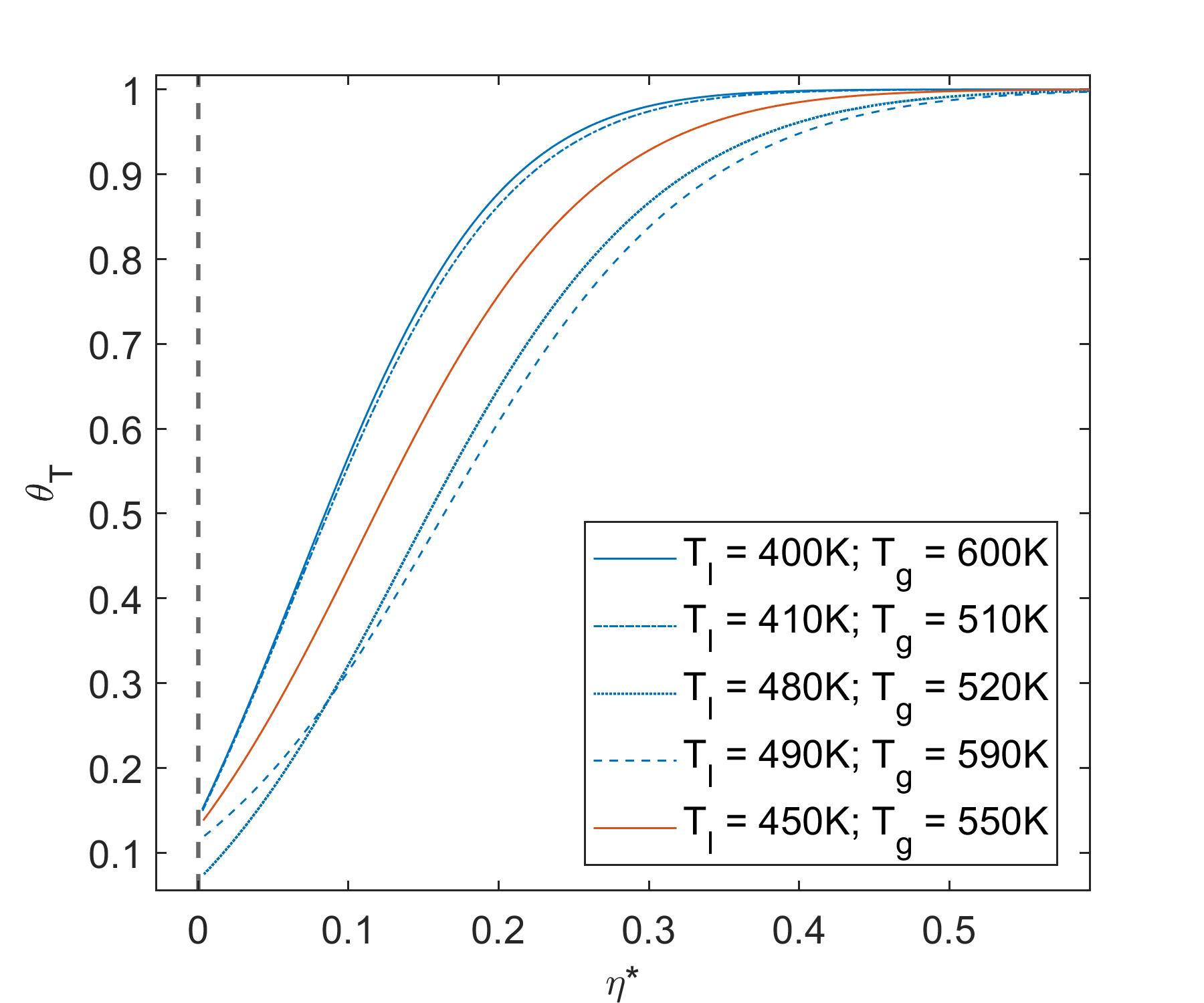}
  \caption{} 
\end{subfigure}%
\begin{subfigure}{.5\textwidth}
  \centering
  \includegraphics[width=1.0\linewidth]{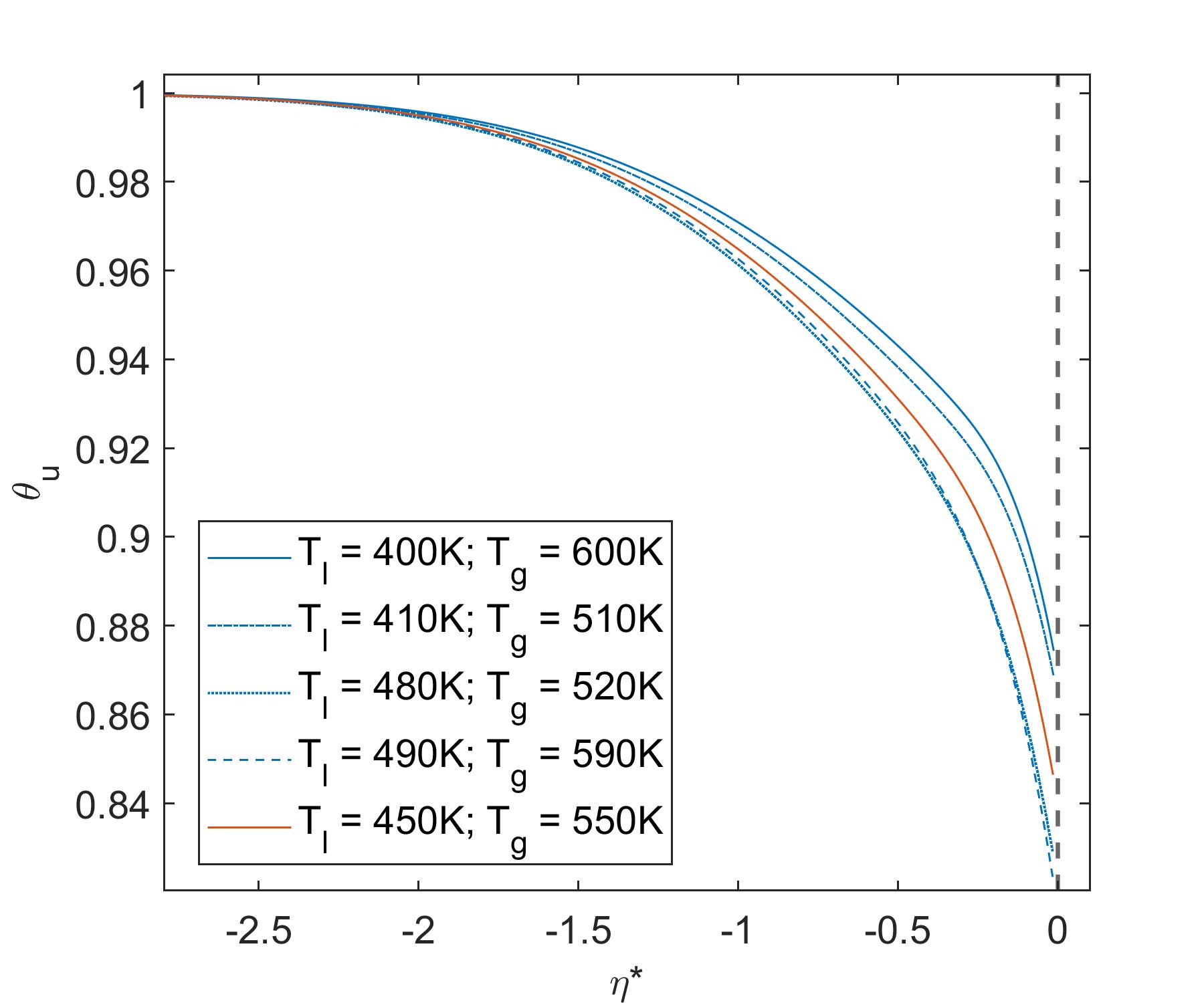}
  \caption{}
\end{subfigure}%
\
\caption{Comparison of the temperature and streamwise velocity distributions in the transverse direction against the similarity variable, \(\eta \) for the oxygen/n-decane mixture at \(p =\) 150 bar and streamwise position \(x = 1\) cm. (a) normalized temperature in the liquid phase near the interface; (b) normalized streamwise velocity in the liquid phase near the interface.}
\label{fig:tempstreamLVarT}
\end{figure}

The same conclusion can be made for varying temperature cases. This can be seen in Figure \ref{fig:tempstreamLVarT}. Changing the bulk temperature conditions for both the liquid and gas phase affects the thermodynamics considerably and thus, similarity cannot be found. While the streamwise velocity profiles are only weakly dependent on temperature, the interface conditions change causing a non-similar behavior for varying temperature ranges. Profiles for cases with comparable freestream liquid temperatures show stronger congruity regardless of the bulk gas temperatures.

\subsection{Boundary layer approximation}
\label{subsec:simil}

The stress tensor can be written in a generalized form for a Newtonian fluid as

\begin{equation}
\label{eqn:stresstensor}
\tau_{ij} = \mu \bigg{(} \frac{\partial u_{i}}{\partial x_{j}}+ \frac{\partial u_{j}}{\partial x_{i}} \bigg{)} + \delta_{ij} \lambda\frac{\partial u_{k}}{\partial x_{k}}
\end{equation}
\noindent
where \(\tau\) is the deviatoric of the deformation rate tensor, \(\delta\) is the Kronecker delta, and \(\mu\) and \(\lambda\) are the dynamic viscosiy and second coefficient of viscosity. The Stokes' hypothesis is used, whereby \(\lambda = - \frac{2}{3}\mu\).

Differentiating \(\tau\) along the streamwise face with respect to the streamwise and transverse directions yields

\begin{equation}
\label{eqn:tauxx}
\frac{\partial}{\partial x}(\tau_{xx})= 2\frac{\partial}{\partial x}\bigg{(}\mu\frac{\partial u}{\partial x}\bigg{)}-\frac{2}{3}\frac{\partial}{\partial x}\bigg{[}\mu\bigg{(}\frac{\partial u}{\partial x}+\frac{\partial v}{\partial y}\bigg{)}\bigg{]}
\end{equation}

\begin{equation}
\label{eqn:tauxy}
\frac{\partial}{\partial y} (\tau_{xy})= \frac{\partial}{\partial y} \bigg{(}\mu \frac{\partial u}{\partial y}\bigg{)}+\frac{\partial}{\partial y} \bigg{(}\mu \frac{\partial v}{\partial x}\bigg{)}
\end{equation}

\begin{figure}[h!]
\includegraphics[scale=.47]{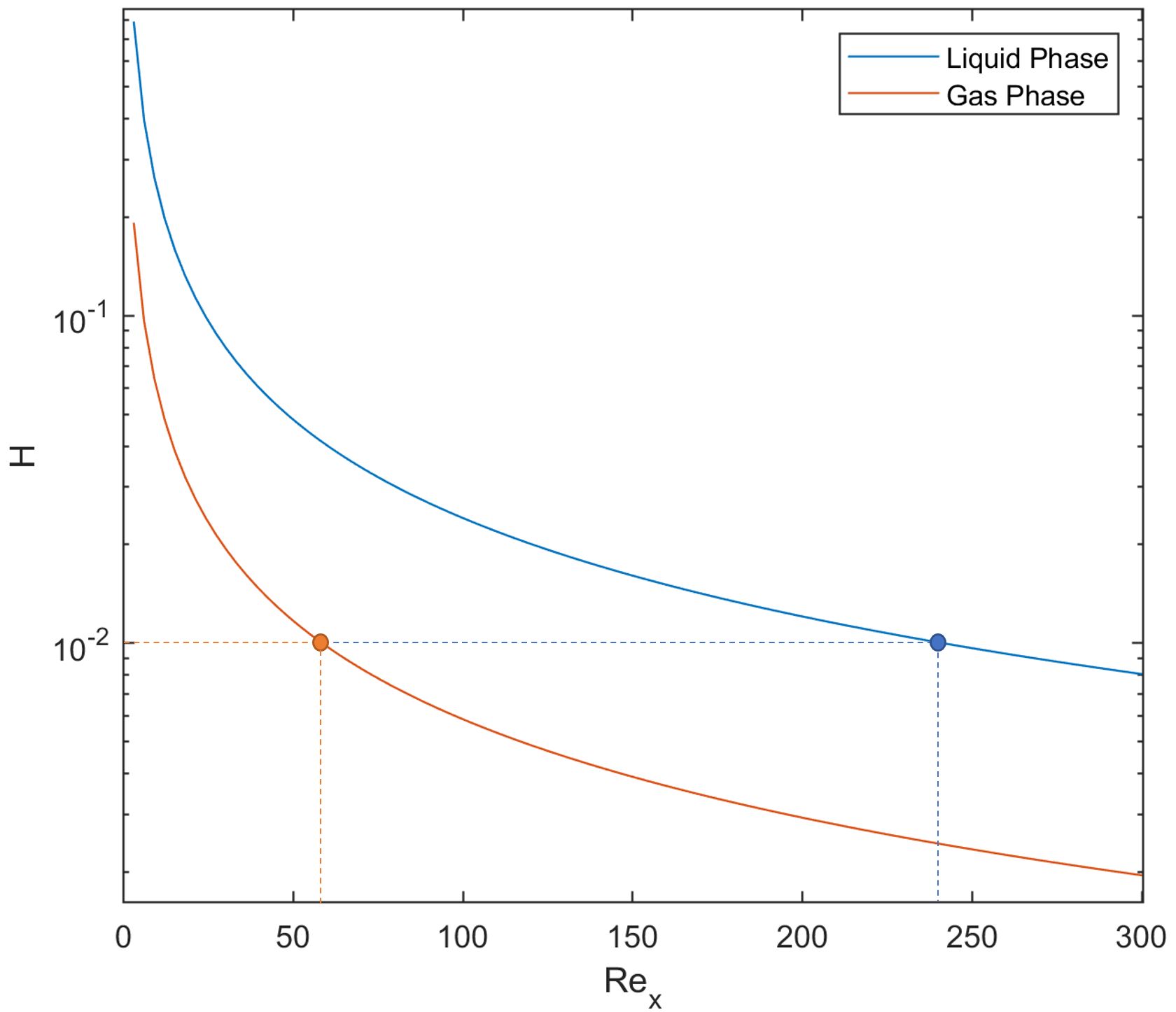}
\centering
\caption{\(H\) for varying streamwise Reynolds numbers at \(p =\) 150 bar.}
\label{fig:ShearTermRatios}
\end{figure}

A boundary layer approximation is assumed in this work. Thus, normal stress, \(\tau_{xx}\), and the streamwise derivative term in \(\tau_{xy}\) are neglected. \(H\) is defined as the ratio of terms not included to those included as

\begin{equation}
H = \frac{\frac{\partial}{\partial x}\bigg{(}\mu\frac{\partial u}{\partial x}\bigg{)} + \frac{\partial}{\partial x}\bigg{[}\mu\bigg{(}\frac{\partial u}{\partial x}+\frac{\partial v}{\partial y}\bigg{)}\bigg{]} + \frac{\partial}{\partial y} \bigg{(}\mu \frac{\partial v}{\partial x}\bigg{)}}{\frac{\partial}{\partial y} \bigg{(}\mu \frac{\partial u}{\partial y}\bigg{)}}
\label{eq:ShearTermRatio}
\end{equation}
\noindent
Similarity can only be believed in the region where the mixing layer equations are valid. \(H < 10^{-2}\) is considered sufficient for the boundary layer approximation to hold. The validity of the mixing layer equations is shown in Figure \ref{fig:ShearTermRatios}, which plots \(H\) for varying streamwise Reynolds numbers at 150 bar with a semilog scale. For both liquid and gas phases, \(H\) decreases as the Reynolds number is increased. The effects of \(\tau_{xx}\) taper off with streamwise distance. Immediately at a unity Reynolds number, \(\frac{\partial}{\partial y} \bigg{(}\mu \frac{\partial u}{\partial y}\bigg{)}\) is greater than all other terms combined for both phases. Progressing downstream, \(\frac{\partial}{\partial y} \bigg{(}\mu \frac{\partial u}{\partial y}\bigg{)}\) becomes two orders of magnitude larger than the other terms at \(Re_{x} = 57\) and \(Re_{x} = 239\) in the gas and liquid phases respectively. This suggests that the similarity analysis performed in this work is valid for streamwise Reynolds numbers greater than 239 at 150 bar. That is, for \(Re_x > 239\), terms such as \(\tau_{xx}\) and the streamwise derivatives in \(\tau_{xy}\) can be safely neglected when considering the one-dimensional profile similarity. 

An analysis for the boundary layer approximation validity has not been evaluated for 10, 50, and 100 bar. However, figures shown in Section \ref{subsec:simil} present profiles at very high Reynolds numbers. Thereby, the results can confidently be believed for all pressure cases.

\section{Summary and conclusions}
\label{summary}

The variable-density, multicomponent laminar boundary-layer equations coupled with a real-fluid thermodynamic model were used to analyze the resulting shear layer between a cold liquid and hot gas. They show that a sharp phase interface still exists at pressures above the critical pressure for both the liquid \(n\)-decane and gaseous oxygen.

A Kelvin-Helmholtz instability analysis was performed to ensure continuum behavior and to show phase equilibrium could be established in a shorter distance than required for amplitude growth with hydrodynamic instabilities. A transverse velocity was found to balance the transverse momentum on either side of the interface. However, interface deflection was between 2 to 3 orders of magnitude smaller than the diffusion layer thickness. Thus, for all pressure cases, a fixed interface at \(y\) = 0 is reasonable. 

Thick diffusion layers form around the liquid-gas interface at \(x =\) 1 cm downstream for different pressures (i.e., 10 - 18 \(\mu\)m in the liquid phase and 30 - 160 \(\mu\)m in the gas phase). It was found that the diffusion layer thickness increased in the liquid phase and decreased in the gas phase as pressure increased. While gaseous oxygen dissolves and liquid \(n\)-decane mixes with the gas vaporizes at all pressures, a transition from net vaporization to condensation of the liquid phase occurred around 50 bar. 

Reduction to a form was found where the dependent variable profiles at different downstream positions collapse onto each other when plotted verses the similarity variable, \(\eta\). When pressure or temperature is varied, similarity for differing constraints is lost as the thermodynamics change the interface properties and the diffusion layer evolution. However, the interface temperature is strongly dependent on the bulk liquid phase temperature allowing for stronger similarity between similar bulk liquid temperatures regardless of the bulk gas temperature. Similarity under the boundary layer approximation was shown to hold in both liquid and gas phases for Reynolds numbers greater than 239 at 150 bar. Profiles presented are at very high Reynolds numbers. Therefore, the results are expected for all analyzed pressures.

\section*{Conflict of interest}

The authors declared that there is no conflict of interest.

\section*{Acknowledgments}

The authors are grateful for the support of the
NSF grant with Award Number 1803833 and Dr.
Ron Joslin as Scientific Officer.



\newpage
\appendix
\section{Evaluation of enthalpies, internal energy, entropy, fugacity coefficient, and specific heat at constant pressure}
\label{apn:thermo1}

Expressions for enthalpy, \(h\), Eq.~(\ref{eqn:enthrel}); internal energy, \(e\), Eq.~(\ref{eqn:intenerel}); entropy, \(s\), Eq.~(\ref{eqn:entrrel}); and fugacity coefficient, \(\Phi\), Eq.~(\ref{eqn:fugrel}), are derived from fundamental thermodynamic principles, where deviations from the ideal gas state (denoted by \(^*\)) are introduced by means of a departure function~\cite{poling2001properties}. \par

\begin{equation}
\label{eqn:enthrel}
h=h^*(T)+\frac{1}{MW}\Bigg[R_uT(Z-1)+R_uT\int_{\infty}^{\bar{v}}\Bigg(T\bigg(\frac{\partial Z}{\partial T}\bigg)_{\bar{v},X_i}\Bigg)\frac{d\bar{v}}{\bar{v}}\Bigg]
\end{equation}

\begin{equation}
\label{eqn:intenerel}
e=e^*(T)+\frac{R_uT}{MW}\int_{\infty}^{\bar{v}}\Bigg(T\bigg(\frac{\partial Z}{\partial T}\bigg)_{\bar{v},X_i}\Bigg)\frac{d\bar{v}}{\bar{v}}
\end{equation}

\begin{equation}
\label{eqn:entrrel}
\begin{split}
s = & s^*(T,p_0)+\frac{1}{MW}\bigg[-R_u\ln\bigg(\frac{p}{p_0}\bigg)-R_u\sum_{i=1}^{N}X_i\ln(X_i)\bigg]\\
& + \frac{1}{MW}\bigg[R_u\ln(Z) + R_u\int_{\infty}^{\bar{v}}\bigg(T\bigg(\frac{\partial Z}{\partial T}\bigg)_{\bar{v},X_i}-1+Z\bigg)\frac{d\bar{v}}{\bar{v}}\bigg]
\end{split}
\end{equation}

\begin{equation}
\label{eqn:fugrel}
\ln(\Phi_i)=\int_{\infty}^{\bar{v}}\Bigg[\frac{1}{\bar{v}}-\frac{1}{R_uT}\Bigg(\frac{\partial p}{\partial X_i}\Bigg)_{T,\bar{v},X_{j\neq i}}\bigg]d\bar{v}-\ln Z
\end{equation}

In Eq.~(\ref{eqn:enthrel})-(\ref{eqn:fugrel}), \(MW\) states the molecular weight of the mixture. Note that Eq.~(\ref{eqn:entrrel}) also includes terms to account for deviations from the reference pressure of the ideal gas mixture entropy and the entropy of mixing caused by the irreversible mixing process between different species~\cite{turns1996introduction,tillner1998helmholtz,neto2013departure}. Combining the previous expressions with the modified SRK equation of state, it yields

\begin{equation}
\label{eqn:enthSRKmod}
\begin{split}
h=h^*(T)&+\frac{1}{MW}\Bigg[R_uT(Z-1)+\frac{T(\partial a/\partial T)_{\bar{v},X_i}-a}{b}\ln\bigg(1+\frac{B}{Z+C}\bigg)
+R_uT^2\bigg(\frac{Z}{\bar{v}}\bigg)\bigg(\frac{\partial c}{\partial T}\bigg)_{\bar{v},X_i}\Bigg]
\end{split}
\end{equation}

\begin{equation}
\label{eqn:inteneSRKmod}
e=e^*(T)+\frac{1}{MW}\Bigg[\frac{T(\partial a/\partial T)_{\bar{v},X_i}-a}{b}\ln\bigg(1+\frac{B}{Z+C}\bigg)+R_uT^2\bigg(\frac{Z}{\bar{v}}\bigg)\bigg(\frac{\partial c}{\partial T}\bigg)_{\bar{v},X_i}\Bigg]
\end{equation}

\begin{equation}
\label{eqn:entrSRKmod}
\begin{split}
s = & s^*(T,p_0)+\frac{1}{MW}\bigg[-R_u\ln\bigg(\frac{p}{p_0}\bigg)-R_u\sum_{i=1}^{N}X_i\ln(X_i)\bigg]\\
& + \frac{1}{MW}\bigg[ \frac{1}{b}\bigg(\frac{\partial a}{\partial T}\bigg)_{\bar{v},X_i}\ln\bigg(1+\frac{B}{Z+C}\bigg) + R_u\ln(Z+C-B)\bigg] \\
& + \frac{1}{MW}\bigg[ R_uT\bigg(\frac{Z}{\bar{v}}\bigg)\bigg(\frac{\partial c}{\partial T}\bigg)_{\bar{v},X_i} \bigg]
\end{split}
\end{equation}

\begin{equation}
\label{eqn:fugSRKmod1}
\ln(\Phi_i) = \frac{Z+C-1}{b}\frac{\partial b}{\partial X_i}-\frac{C}{c}\frac{\partial c}{\partial X_i}-\ln(Z+C-B)-\frac{A}{B}\bigg(\frac{1}{a}\frac{\partial a}{\partial X_i}-\frac{1}{b}\frac{\partial b}{\partial X_i}\bigg)\ln\bigg(1+\frac{B}{Z+C}\bigg)
\end{equation}

Furthermore, expressions for the specific heat at constant pressure and partial enthalpy of species \(i\) in a mixture are needed. They are obtained by applying the respective thermodynamic definitions. The specific heat at constant pressure, \(c_p\), becomes

\begin{equation}
\begin{split}
c_p = \bigg(\frac{\partial h}{\partial T}\bigg)_{p,X_i} & = c_{p}^{*}(T)+\frac{1}{MW}\Bigg[\frac{T}{b}\bigg(\frac{\partial^2 a}{\partial T^2}\bigg)_{p,X_i}\ln\bigg(1+\frac{B}{Z+C}\bigg)-R_u\Bigg] \\
& + \frac{1}{MW}\Bigg[\bigg(p-\frac{T(\partial a/\partial T)_{\bar{v},X_i} - a}{(\bar{v}+c)(\bar{v}+c+b)}\bigg)\bigg(\bigg(\frac{\partial \bar{v}}{\partial T}\bigg)_{p,X_i}+\bigg(\frac{\partial c}{\partial T}\bigg)_{p,X_i}\bigg)\Bigg] \\
& + \frac{1}{MW}\Bigg[R_uT^2\bigg(\frac{Z}{\bar{v}}\bigg)\bigg(\frac{\partial^2 c}{\partial T^2}\bigg)_{p,X_i}\Bigg]
\end{split}
\end{equation}

\noindent
and the partial molar enthalpy is

\begin{equation}
\begin{split}
\bar{h}_i & = \bigg(\frac{\partial \bar{h}}{\partial X_i}\bigg)_{p,T,X_{j\neq i}} = \bar{h}_{i}^{*}(T)+p\bigg(\frac{\partial \bar{v}}{\partial X_i}\bigg)_{p,T,X_{j\neq i}}-R_uT \\
&+ \frac{aA_1}{\bar{v}+c+b}\Bigg[A_2-\frac{1}{\bar{v}+c}\Bigg(\bigg(\frac{\partial \bar{v}}{\partial X_i}\bigg)_{p,T,X_{j\neq i}}+\bigg(\frac{\partial c}{\partial X_i}\bigg)_{p,T,X_{j\neq i}}\Bigg)\Bigg]\\
&+\frac{1}{b}\Bigg(T\bigg(\frac{\partial^2 a}{\partial X_i \partial T}\bigg)_{p,T,X_{j\neq i}}-\bigg(\frac{\partial a}{\partial X_i}\bigg)_{p,T,X_{j\neq i}}-aA_1A_2\Bigg)\ln\bigg(\frac{\bar{v}+c+b}{\bar{v}+c}\bigg) \\
& + R_uT^2\bigg(\frac{Z}{\bar{v}}\bigg)\bigg(\frac{\partial^2 c}{\partial X_i \partial T}\bigg)_{p,T,X_{j\neq i}}
\end{split}
\end{equation}

\noindent 
where \(A_1\) and \(A_2\) are defined as

\begin{equation}
A_1 \equiv \frac{T}{a}\bigg(\frac{\partial a}{\partial T}\bigg)_{v,X_i}-1 \quad ; \quad A_2 \equiv \frac{1}{b}\bigg(\frac{\partial b}{\partial X_i}\bigg)_{v,X_i}
\end{equation}

Once the partial molar enthalpy is known, the partial specific enthalpy needed in Eq.~(\ref{eqn:ene_noncons}) is obtained as \(h_i = \bar{h}_i/MW_i\). \par 

All partial derivatives involved in the previous expressions can be found in \ref{apn:thermo}. Ideal-gas enthalpy, internal energy, entropy and specific heat at constant pressure are obtained from the correlations by Passut and Danner~\cite{passut1972correlation} and ideal gas mixing rules as

\begin{equation}
h^*(T) = \sum_{i=1}^{N}Y_ih_{i}^{*}(T) \quad ; \quad e^*(T) = h^*(T) - p/\rho^* 
\end{equation}

\begin{equation}
s^*(T,p_0) = \sum_{i=1}^{N}Y_is_{i}^{*}(T,p_0) \quad ; \quad c_{p}^{*}(T) = \sum_{i=1}^{N}Y_ic_{p_i}^{*}(T)
\end{equation}

\noindent
with

\begin{equation}
\label{eqn:idealenth}
h_{i}^{*}(T)=\hat{A} + \hat{B}T + \hat{C}T^2 + \hat{D}T^3 + \hat{E}T^4 + \hat{F}T^5
\end{equation}

\begin{equation}
\label{eqn:idealentr}
s_{i}^{*}(T,p_0)=\hat{B}\ln T + 2\hat{C}T + \frac{3}{2}\hat{D}T^2 + \frac{4}{3}\hat{E}T^3 + \frac{5}{4}\hat{F}T^4 + \hat{G}
\end{equation}

\begin{equation}
\label{eqn:idealcp}
c_{p,i}^{*}(T)=\hat{B} + 2\hat{C}T + 3\hat{D}T^2 + 4\hat{E}T^3 + 5\hat{F}T^4
\end{equation}

\noindent
where \(\rho^*\) is the ideal gas density evaluated using the ideal gas law, \(p_0\) is the reference pressure for entropy calculations set at 1 atm and \(\hat{A}\)-\(\hat{G}\) are correlation constants given in~\cite{passut1972correlation}.

\newpage
\section{Thermodynamic derivatives based on the SRK equation of state}
\label{apn:thermo}

\subsection{``a" derivatives}

\begin{equation}
\bigg(\frac{\partial a}{\partial X_i}\bigg)_{T,\bar{v},X_{j\neq i}}=2\sum_{j=1}^{N}X_j(a_ia_j)^{0.5}(1-k_{ij}) 
\end{equation}

\begin{equation}
\bigg(\frac{\partial^2 a}{\partial X_j \partial X_i}\bigg)_{p,T} = 2(a_ia_j)^{0.5}(1-k_{ij})
\end{equation}

\begin{equation}
\bigg(\frac{\partial^2 a}{\partial X_i \partial T}\bigg)_{p,T,X_{j\neq i}} = \sum_{J=1}^{N} X_j \Bigg[\bigg(\frac{a_i}{a_j}\bigg)^{0.5}\frac{da_j}{dT}+\bigg(\frac{a_j}{a_i}\bigg)^{0.5}\frac{da_i}{dT}\Bigg](1-k_{ij})
\end{equation}

\begin{equation}
\bigg(\frac{\partial a}{\partial T}\bigg)_{\bar{v},X_i} = \frac{1}{2}\sum_{i=1}^{N}\sum_{j=1}^{N} X_iX_j\Bigg[\bigg(\frac{a_i}{a_j}\bigg)^{0.5}\frac{da_j}{dT}+\bigg(\frac{a_j}{a_i}\bigg)^{0.5}\frac{da_i}{dT}\Bigg](1-k_{ij})
\end{equation}

\begin{equation}
\begin{split}
\bigg(\frac{\partial^2 a}{\partial T^2}\bigg)_{p,X_i} & = \frac{1}{2}\sum_{i=1}^{N}\sum_{j=1}^{N}X_iX_j\Bigg[\bigg(\frac{a_i}{a_j}\bigg)^{0.5}\frac{d^2a_j}{dT^2}+\bigg(\frac{a_j}{a_i}\bigg)^{0.5}\frac{d^2a_i}{dT^2}+\bigg(\frac{1}{a_ia_j}\bigg)^{0.5}\frac{da_i}{dT}\frac{da_j}{dT}\Bigg](1-k_{ij}) \\
 & - \frac{1}{4}\sum_{i=1}^{N}\sum_{j=1}^{N}X_iX_j\Bigg[\bigg(\frac{a_i}{a_j}\bigg)^{0.5}\frac{1}{a_j}\bigg(\frac{da_j}{dT}\bigg)^2+\bigg(\frac{a_j}{a_i}\bigg)^{0.5}\frac{1}{a_i}\bigg(\frac{da_i}{dT}\bigg)^2\Bigg](1-k_{ij})
 \end{split}
\end{equation}

\begin{equation}
\frac{da_i}{dT}=a_{c_i}\frac{d\alpha_i}{dT}=\frac{a_i}{\alpha_i}\frac{d\alpha_i}{dT}
\end{equation}

\begin{equation}
\frac{d^2a_i}{dT^2}=a_{c_i}\frac{d^2\alpha_i}{dT^2}=\frac{a_i}{\alpha_i}\frac{d^2\alpha_i}{dT^2}
\end{equation}

\begin{equation}
\frac{d^2\alpha_i}{dT^2} = \frac{1}{2}\Bigg[\frac{1}{\alpha_i}\bigg(\frac{d\alpha_i}{dT}\bigg)^2-\frac{1}{T}\frac{d\alpha_i}{dT}\Bigg]
\end{equation}

\subsection{``b" derivatives}

\begin{equation}
\bigg(\frac{\partial b}{\partial X_i}\bigg)_{T,\bar{v},X_{j\neq i}} = b_i
\end{equation}

\subsection{``c" derivatives}

\begin{equation}
\bigg(\frac{\partial c}{\partial X_i}\bigg)_{T,\bar{v},X_{j\neq i}}=c_i
\end{equation}

\begin{equation}
\bigg(\frac{\partial^2 c}{\partial X_i \partial T}\bigg)_{p,T,X_{j\neq i}} = \bigg(\frac{\partial c_i}{\partial T}\bigg)_{\bar{v},X_i} = \frac{c_{c_i}}{T_{c_i}}f'(T_{r_i})
\end{equation}

\begin{equation}
\bigg(\frac{\partial c}{\partial T}\bigg)_{\bar{v},X_i} = \sum_{i=1}^{N} X_i \bigg(\frac{\partial c_i}{\partial T}\bigg)_{\bar{v},X_i}= \sum_{i=1}^{N} X_i \frac{c_{c_i}}{T_{c_i}}f'(T_{r_i})
\end{equation}

\begin{align}
\bigg(\frac{\partial^2 c}{\partial T^2}\bigg)_{p,X_i} = \sum_{i=1}^{N} X_i \bigg(\frac{\partial^2 c_i}{\partial T^2}\bigg)_{\bar{v},X_i}= \sum_{i=1}^{N} X_i \frac{c_{c_i}}{T_{c_i}^{2}}f''(T_{r_i})
\end{align}

\subsection{``\(\bar{v}\)" derivatives}

\begin{equation}
\begin{split}
\bigg(\frac{\partial \bar{v}}{\partial X_i}\bigg)_{p,T,X_{j\neq i}} & = \frac{R_uT}{p}\Bigg(Z+C-\frac{C}{c}\bigg(\frac{\partial c}{\partial X_i}\bigg)_{p,T,X_{j\neq i}}\Bigg) \\
&+ \frac{R_uT}{p}\Bigg(\frac{(B-Z-C)\big[\frac{A}{a}\big(\frac{\partial a}{\partial X_i}\big)_{p,T,X_{j\neq i}}-2A\big]}{3Z^2+2(3C-1)Z+3C^2-2C+A-B-B^2}\Bigg) \\
&+ \frac{R_uT}{p}\Bigg(\frac{(Z+C+2B[Z+C]+A)\big[\frac{B}{b}\big(\frac{\partial b}{\partial X_i}\big)_{p,T,X_{j\neq i}}-B\big]}{3Z^2+2(3C-1)Z+3C^2-2C+A-B-B^2}\Bigg)
\end{split}
\end{equation}

\begin{equation}
\begin{split}
\bigg(\frac{\partial \bar{v}}{\partial T}\bigg)_{p,X_i} & = -\bigg(\frac{\partial c}{\partial T}\bigg)_{p,X_i}+\frac{R_uT}{p}\Bigg(\frac{Z+C}{T}\Bigg) \\
&-\frac{R_uT}{p}\Bigg(\frac{(Z+C-B)\big(\frac{A}{a}(\frac{\partial a}{\partial T})_{p,X_i}-2\frac{A}{T}\big)+(Z+C+2B[Z+C]+A)\frac{B}{T}}{3Z^2+2(3C-1)Z+3C^2-2C+A-B-B^2}\Bigg)
\end{split}
\end{equation}

\subsection{``\(\Phi\)" derivatives}

\begin{equation}
\begin{split}
\bigg(\frac{\partial \ln(\Phi_i)}{\partial X_j}\bigg)_{p,T,X_{i\neq j}} & = \frac{B}{Z+C-B}\frac{1}{b}\frac{\partial b}{\partial X_j}-\frac{Z+C-1}{b^2}\frac{\partial b}{\partial X_i}\frac{\partial b}{\partial X_j} \\
& -(Z+C)\frac{1}{b}\frac{\partial b}{\partial X_i} + \frac{C}{c}\frac{\partial c}{\partial X_i} + 1 \\
& + \frac{A}{Z+C+B}\frac{1}{b^2}\frac{\partial b}{\partial X_i}\frac{\partial b}{\partial X_j}-\frac{A}{Z+C+B}\frac{1}{ab}\frac{\partial a}{\partial X_i}\frac{\partial b}{\partial X_j} \\
& +\frac{A}{B}\Bigg(\frac{1}{ab}\frac{\partial a}{\partial X_i}\frac{\partial b}{\partial X_j}+\frac{1}{ab}\frac{\partial a}{\partial X_j}\frac{\partial b}{\partial X_i}\Bigg)\ln\bigg(1+\frac{B}{Z+C}\bigg) \\
& - \frac{A}{B}\Bigg(\frac{2}{b^2}\frac{\partial b}{\partial X_i}\frac{\partial b}{\partial X_j}+\frac{1}{a}\frac{\partial^2 a}{\partial X_j \partial X_i}\Bigg)\ln\bigg(1+\frac{B}{Z+C}\bigg) \\
& +\frac{A}{B}\Bigg(\frac{1}{a}\frac{\partial a}{\partial X_i}-\frac{1}{b}\frac{\partial b}{\partial X_i}\Bigg)\ln\bigg(1+\frac{B}{Z+C}\bigg) \\
& + \Bigg(\frac{1}{b}\frac{\partial b}{\partial X_i}-\frac{1}{Z+C-B}\Bigg)\Bigg(\frac{Z}{\bar{v}}\frac{\partial \bar{v}}{\partial X_j}+\frac{C}{c}\frac{\partial c}{\partial X_j}\Bigg) \\
& + \frac{A}{Z+C+B}\Bigg(\frac{1}{a}\frac{\partial a}{\partial X_i}-\frac{1}{b}\frac{\partial b}{\partial X_i}\Bigg)\frac{1}{\bar{v}+c}\Bigg(\frac{\partial \bar{v}}{\partial X_j}+\frac{\partial c}{\partial X_j}\Bigg) 
\end{split}
\end{equation}

\begin{equation}
\begin{split}
\bigg(\frac{\partial \ln(\Phi_i)}{\partial T}\bigg)_{p,X_i} & = \Bigg(1-\frac{Z+C}{b}\frac{\partial b}{\partial X_i}\Bigg)\frac{1}{T} + \frac{C}{c}\Bigg(\frac{1}{T}\frac{\partial c}{\partial X_i}-\frac{\partial^2 c}{\partial T \partial X_i}\Bigg) \\
& + \frac{A}{B}\Bigg(\frac{1}{a^2}\frac{\partial a}{\partial T}\frac{\partial a}{\partial X_i}-\frac{1}{a}\frac{\partial^2 a}{\partial T \partial X_i}\Bigg)\ln\bigg(1+\frac{B}{Z+C}\bigg) \\
& + \frac{A}{B}\Bigg(\frac{1}{T}-\frac{1}{a}\frac{\partial a}{\partial T}\Bigg)\Bigg(\frac{1}{a}\frac{\partial a}{\partial X_i}-\frac{1}{b}\frac{\partial b}{\partial X_i}\Bigg)\ln\bigg(1+\frac{B}{Z+C}\bigg) \\
& + \Bigg(\frac{1}{b}\frac{\partial b}{\partial X_i}-\frac{1}{Z+C-B}\Bigg)\Bigg(\frac{Z}{\bar{v}}\frac{\partial \bar{v}}{\partial T}+\frac{C}{c}\frac{\partial c}{\partial T}\Bigg) \\
&  + \frac{A}{Z+C+B}\Bigg(\frac{1}{a}\frac{\partial a}{\partial X_i}-\frac{1}{b}\frac{\partial b}{\partial X_i}\Bigg)\frac{1}{\bar{v}+c}\Bigg(\frac{\partial \bar{v}}{\partial T}+\frac{\partial c}{\partial T}\Bigg)
\end{split}
\end{equation}


\newpage
\bibliography{paper}

\begin{thebibliography}{10}
\expandafter\ifx\csname url\endcsname\relax
  \def\url#1{\texttt{#1}}\fi
\expandafter\ifx\csname urlprefix\endcsname\relax\def\urlprefix{URL }\fi
\expandafter\ifx\csname href\endcsname\relax
  \def\href#1#2{#2} \def\path#1{#1}\fi

\bibitem{chapman1949laminar}
D.~Chapman, Laminar mixing of a compressible fluid, NACA Technical Report 958
  (1950) 231--237.

\bibitem{doughty1992similarity}
C.~Doughty, K.~Pruess, A similarity solution for two-phase water, air, and heat
  flow near a linear heat source in a porous medium, Journal of Geophysical
  Research 97 (1992) 1821--1838.

\bibitem{sadatomi1982twophase}
M.~Sadatomi, Y.~Sato, Two-phase flow in vertical noncircular channels,
  International Journal of Multiphase Flow 8 (1982) 641--655.

\bibitem{kleinstreuer2003two}
C.~Kleinstreuer, Two-phase flow: theory and applications, CRC Press, 2003.

\bibitem{white2006viscous}
F.~M. White, I.~Corfield, Viscous {F}luid {F}low, Vol.~3, McGraw-Hill New York,
  2006.

\bibitem{williams2018combustion}
F.~A. Williams, Combustion theory, CRC Press, 2018.

\bibitem{hirschfelder1964molecular}
J.~O. Hirschfelder, C.~F. Curtiss, R.~B. Bird, M.~G. Mayer, Molecular theory of
  gases and liquids, Vol. 165, Wiley New York, 1964.

\bibitem{prausnitz1998molecular}
J.~M. Prausnitz, R.~N. Lichtenthaler, E.~G. De~Azevedo, Molecular
  thermodynamics of fluid-phase equilibria, Pearson Education, 1998.

\bibitem{chehroudi1999initial}
B.~Chehroudi, D.~Talley, E.~Coy, Initial growth rate and visual characteristics
  of a round jet into a sub-to supercritical environment of relevance to
  rocket, gas turbine, and diesel engines, 37th Aerospace Sciences Meeting and
  Exhibit (1999) 1999--206.

\bibitem{maslowe1971inviscid}
S.~Maslowe, R.~Kelly, Inviscid instability of an unbounded heterogeneous shear
  layer, Journal of Fluid Mechanics 48~(2) (1971) 405--415.

\bibitem{mayer1998propellant}
W.~Mayer, B.~Ivancic, A.~Schik, U.~Hornung, Propellant atomization in
  {LOX}/{GH2} rocket engines, 34th AIAA/ASME/SAE/ASEE Joint Propulsion
  Conference and Exhibit (1998) 1998--3685.

\bibitem{hsieh1991droplet}
K.~C. Hsieh, J.~S. Shuen, V.~Yang, Droplet vaporization in high-pressure
  environments {I}: {N}ear critical conditions, Combustion Science and
  Technology 76~(1-3) (1991) 111--132.

\bibitem{delplanque1993numerical}
J.-P. Delplanque, W.~A. Sirignano, Numerical study of the transient
  vaporization of an oxygen droplet at sub- and supercritical conditions,
  International Journal of Heat and Mass Transfer 36~(2) (1993) 303--314.

\bibitem{yang1994vaporization}
V.~Yang, J.-S. Shuen, Vaporization of liquid oxygen ({LOX}) droplets in
  supercritical hydrogen environments, Combustion Science and Technology
  97~(4-6) (1994) 247--270.

\bibitem{sirignano1997selected}
W.~A. Sirignano, J.-P. Delplanque, F.~Liu, Selected challenges in jet and
  rocket engine combustion research, 33rd Joint Propulsion Conference and
  Exhibit (1997) 1997--2701.

\bibitem{juanos2015thermodynamic}
A.~Jord{\`a}-Juan{\'o}s, W.~A. Sirignano, Thermodynamic analysis for combustion
  at high gas densities, Proceedings of the 25th ICDERS. Leeds, UK: ICDERS.

\bibitem{mayer1996propellant}
W.~Mayer, H.~Tamura, Propellant injection in a liquid oxygen/gaseous hydrogen
  rocket engine, Journal or Propulsion and Power 12 (1996) 1137--1147.

\bibitem{mayer2000injection}
W.~Mayer, A.~Schik, M.~Schaffler, H.~Tamura, Injection and mixing processes in
  high-pressure liquid oxygen/gaseous hydrogen rocket combustors, Journal of
  Propulsion and Power 16 (2000) 823--828.

\bibitem{segal2008subcritical}
C.~Segal, A.~Polikhov, Subcritical to supercritical mixing, Physics of Fluids
  20 (2008) 052101.

\bibitem{chehroudi2012recent}
B.~Chehroudi, Recent experimental efforts on high-pressure supercritical
  injection for liquid rockets and their implications, International Journal of
  Aerospace Engineering 2012 (2012) 121802.

\bibitem{spalding1959theory}
D.~Spalding, Theory of particle combustion at high pressures, ARS journal
  29~(11) (1959) 828--835.

\bibitem{rosner1967liquid}
D.~Rosner, On liquid droplet combustion at high pressures, AIAA Journal 5~(1)
  (1967) 163--166.

\bibitem{poblador2018transient}
J.~Poblador-Ibanez, W.~A. Sirignano, Transient behavior near liquid-gas
  interface at supercritical pressure, International Journal of Heat and Mass
  Transfer 126 (2018) 457--473.

\bibitem{poblador2020self}
J.~Poblador-Ibanez, B.~Davis, W.~Sirignano, Self-similar solution of a
  supercritical two-phase laminar mixing layer, International Journal of
  Multiphase Flow (in review, available at arXiv preprint arXiv:2004.00564
  (2020) 1--44.

\bibitem{poblador2019analysis}
J.~Poblador-Ibandez, W.~A. Sirignano, Analysis of an axisymmetric liquid jet at
  supercritical pressures, ILASS-Americas 30th Annual Conference on Liquid
  Atomization and Spray Systems.

\bibitem{jarrahbashi2014vorticity}
D.~Jarrahbashi, W.~A. Sirignano, Vorticity dynamics for transient high-pressure
  liquid injection, Physics of Fluids 26 (2014) 101304.

\bibitem{jarrahbashi2016early}
D.~Jarrahbashi, W.~A. Sirignano, P.~Popov, F.~Hussain, Early spray development
  at high gas density: hole, ligament and bridge formations, Journal of Fluid
  Mechanics 792 (2016) 186--231.

\bibitem{zandian2017planar}
A.~Zandian, W.~A. Sirignano, F.~Hussain, Planar liquid jet: Early deformation
  and atomization cascades, Physics of Fluids 29 (2017) 062109.

\bibitem{2017arXiv170603742Z}
A.~Zandian, W.~A. Sirignano, F.~Hussain, Understanding liquid-jet atomization
  cascades via vortex dynamics, Journal of Fluid Mechanics 843 (2018) 293--354.

\bibitem{huang1990small-scale}
L.-S. Huang, C.-M. Ho, Small-scale transition in a plane mixing layer, Journal
  of Fluid Mechanics 210 (1990) 475--500.

\bibitem{tani1969boundary}
I.~Tani, Boundary-layer transition, Annual Review of Fluid Mechanics 1 (1969)
  169--196.

\bibitem{rangel1991linear}
R.~Rangel, W.~Sirignano, The linear and nonlinear shear instability of a fluid
  sheet, Physics of Fluids 3 (1991) 2392.

\bibitem{joseph2007potential}
D.~Joseph, Potential {F}lows of {V}iscous and {V}iscoelastic {L}iquids,
  Cambridge {U}niversity {P}ress, 2007.

\bibitem{he2017sharp}
P.~He, A.~F. Ghoniem, A sharp interface method for coupling multiphase flow,
  heat transfer and multicomponent mass transfer with interphase diffusion,
  Journal of Computational Physics 332 (2017) 316--332.

\bibitem{leahy2007unified}
A.~Leahy-Dios, A.~Firoozabadi, Unified model for nonideal multicomponent
  molecular diffusion coefficients, AIChE Journal 53~(11) (2007) 2932--2939.

\bibitem{mutoru2011form}
J.~W. Mutoru, A.~Firoozabadi, Form of multicomponent {F}ickian diffusion
  coefficients matrix, The Journal of Chemical Thermodynamics 43~(8) (2011)
  1192--1203.

\bibitem{soave1972equilibrium}
G.~Soave, Equilibrium constants from a modified {R}edlich-{K}wong equation of
  state, Chemical Engineering Science 27~(6) (1972) 1197--1203.

\bibitem{yang2000modeling}
V.~Yang, Modeling of supercritical vaporization, mixing, and combustion
  processes in liquid-fueled propulsion systems, Proceedings of the Combustion
  Institute 28~(1) (2000) 925--942.

\bibitem{prausnitz2004thermodynamics}
J.~M. Prausnitz, F.~W. Tavares, Thermodynamics of fluid-phase equilibria for
  standard chemical engineering operations, AIChE journal 50~(4) (2004)
  739--761.

\bibitem{lin2006volumetric}
H.~Lin, Y.-Y. Duan, T.~Zhang, Z.-M. Huang, Volumetric {P}roperty {I}mprovement
  for the {S}oave- {R}edlich- {K}wong {E}quation of {S}tate, Industrial \&
  Engineering Chemistry Research 45~(5) (2006) 1829--1839.

\bibitem{graboski1978modified}
M.~S. Graboski, T.~E. Daubert, A modified {S}oave equation of state for phase
  equilibrium calculations. 1. {H}ydrocarbon systems, Industrial \& Engineering
  Chemistry Process Design and Development 17~(4) (1978) 443--448.

\bibitem{graboski1978modified2}
M.~S. Graboski, T.~E. Daubert, A modified {S}oave equation of state for phase
  equilibrium calculations. 2. {S}ystems containing {CO}2, {H2S}, {N2}, and
  {CO}, Industrial \& Engineering Chemistry Process Design and Development
  17~(4) (1978) 448--454.

\bibitem{chung1988generalized}
T.~H. Chung, M.~Ajlan, L.~L. Lee, K.~E. Starling, Generalized multiparameter
  correlation for nonpolar and polar fluid transport properties, Industrial \&
  Engineering Chemistry Research 27~(4) (1988) 671--679.

\bibitem{poling2001properties}
B.~E. Poling, J.~M. Prausnitz, O.~John~Paul, R.~C. Reid, The {P}roperties of
  {G}ases and {L}iquids, Vol.~5, Mc{G}raw-{H}ill New York, 2001.

\bibitem{krishna2016describing}
R.~Krishna, J.~M. van Baten, Describing diffusion in fluid mixtures at elevated
  pressures by combining the maxwell--stefan formulation with an equation of
  state, Chemical Engineering Science 153 (2016) 174--187.

\bibitem{dahms2013transition}
R.~N. Dahms, J.~C. Oefelein, On the transition between two-phase and
  single-phase interface dynamics in multicomponent fluids at supercritical
  pressures, Physics of Fluids 25~(9) (2013) 092103.

\bibitem{dahms2015liquid}
R.~N. Dahms, J.~C. Oefelein, Liquid jet breakup regimes at supercritical
  pressures, Combustion and Flame 162~(10) (2015) 3648--3657.

\bibitem{courant1928partiellen}
R.~Courant, K.~Friedrichs, H.~Lewy, {\"U}ber die partiellen
  differenzengleichungen der mathematischen physik, Mathematische Annalen
  100~(1) (1928) 32--74.

\bibitem{turns1996introduction}
S.~R. Turns, An {I}ntroduction to {C}ombustion, Vol. 499, Mc{G}raw-{H}ill,
  1996.

\bibitem{tillner1998helmholtz}
R.~Tillner-Roth, D.~G. Friend, A {H}elmholtz free energy formulation of the
  thermodynamic properties of the mixture $\{$water+ ammonia$\}$, Journal of
  Physical and Chemical Reference Data 27~(1) (1998) 63--96.

\bibitem{neto2013departure}
M.~A.~M. Neto, J.~R. Barbosa~Jr, A departure-function approach to calculate
  thermodynamic properties of refrigerant-oil mixtures, International Journal
  of Refrigeration 36~(3) (2013) 972--979.

\bibitem{passut1972correlation}
C.~A. Passut, R.~P. Danner, Correlation of ideal gas enthalpy, heat capacity
  and entropy, Industrial \& Engineering Chemistry Process Design and
  Development 11~(4) (1972) 543--546.

\end{thebibliography}

\end{document}